\documentclass[a4paper,11pt]{article}

\usepackage{amsthm,amsmath,amssymb,amsfonts,pstricks,pst-node,graphics}
\usepackage{texdraw}

\DeclareMathAccent{\wtilde}{\mathord}{largesymbols}{"65}
\DeclareMathAccent{\what}{\mathord}{largesymbols}{"62}

\newcommand\cS{{\mathcal S}}
\newcommand\cE{{\mathcal E}}
\newcommand\cF{{\mathcal F}}
\newcommand\cT{{\mathcal T}}

\newcommand\Z{{\mathbb Z}}
\newcommand\N{{\mathbb N}}
\newcommand\C{{\mathbb C}}

\newcommand\J{{\mathcal I}}
\newcommand\cH{{\mathcal H}}

\newcommand{\gA}{{\mathfrak{A}}}
\newcommand{\gP}{{\mathfrak{P}}}

\newcommand{\gM}{{\mathfrak{M}}}

\newcommand{\gR}{{\mathfrak{R}}}

\def\im{\operatorname{Im}}

\newcommand\bs{{\bf s}}
\newcommand\bt{{\bf t}}

\newtheorem{Rem}{Remark}

\newtheorem{Def}{Definition}
\newtheorem{The}{Theorem}
\newtheorem{Pro}{Proposition}
\newtheorem{Lem}{Lemma}

\newtheorem{Cor}{Corollary}

\makeatletter
\def\wb{\accentset{{\cc@style\underline{\mskip10mu}}}}
\makeatother

\begin{document}
\bibliographystyle{unsrt}
\title{Recursion operators, conservation laws and integrability conditions for difference equations}
\author{Alexander V. Mikhailov$^{\star}$, Jing Ping Wang$ ^\dagger $
and Pavlos Xenitidis $^{\star}$\\
$\star$ Applied Mathematics Department, University of Leeds, UK\\
$\dagger$ School of Mathematics and Statistics, University of Kent, UK }
\date{}
\maketitle

\begin{abstract}
In this paper we make an attempt to give a consistent background and definitions
suitable for the theory of integrable difference equations. We adapt a concept of
recursion operator to difference equations and show that it generates an infinite
sequence of symmetries and canonical conservation laws for a difference equation.
Similar to the case of partial differential equations these canonical densities can
serve as integrability conditions for difference equations. We have found the recursion
operators for the Viallet and all ABS equations.
\end{abstract}

\section{Introduction}

The nature of integrability of partial differential equations is rather well
understood \cite{mr91k:58005,integrability}. It can be related to the existence of
Lax representations, Miura and B{\"a}cklund transformations, bi-linear Hirota
representations, multi-soliton solutions, the Painlev\'e property, bi-Hamiltonian
structures, recursion operators and infinite hierarchies of symmetries and
conservation laws. Although there is not any commonly accepted definition for
integrability, the patterns of equations possessing some of the above listed
properties coincide. In the symmetry approach the existence of an infinite hierarchy
of symmetries and conservation laws is taken as a definition of integrability  and
the most complete results on classification of integrable equations are based on this
property \cite{mr86i:58070, mr89g:58092, mr89e:58062, mr93b:58070, mr99g:35058,
wang98, asy}. The symmetry approach  proved to be effective in the problem of
classification of integrable differential-difference equations \cite{Yami1,Yami}.

The theory of integrable difference (or discrete) equations is a relatively recent,
but very active area of research. It inherited some methods and tools from the theory
of integrable partial differential equations, see for instance \cite{NC,GHRV}. The
property of multidimensional consistency, or consistency around the cube \cite{NW},
has been proposed as the integrability criterion for quadrilateral difference
equations \cite{BobSuris,Nijhoff2002}. All integrable affine-linear quadrilateral
equations (satisfying a few symmetry conditions with respect to reflections of the
three dimensional lattice) were classified by Adler, Bobenko and Suris \cite{ABS},
see also \cite{ABS1}. They produced a list of seven equations (up to point
fractional-linear transformations), which will be referred as the ABS equations.
Unfortunately, the method based on the property of multi-dimensional consistency
cannot be directly applied to a given equation on a lattice. Indeed, to check the
consistency around the cube one needs to know the equations corresponding to the
sides of the cube.

Another characteristic property of an integrable equation is the vanishing of its
algebraic entropy  \cite{BellonViallet}. It has been used by Viallet to identify the
equation
\begin{eqnarray}
Q &:=& a_1 u_{0,0} u_{1,0} u_{0,1} u_{1,1} + a_2 (u_{0,0} u_{1,0} u_{0,1} + u_{1,0} u_{0,1} u_{1,1}
 + u_{0,1} u_{1,1} u_{0,0} + u_{1,1} u_{0,0} u_{1,0}) \nonumber\\
&&  + a_3 (u_{0,0} u_{1,0}+u_{0,1} u_{1,1}) +a_4 (u_{1,0} u_{0,1} + u_{0,0} u_{1,1})  + a_5 (u_{0,0} u_{0,1} + u_{1,0} u_{1,1}) \nonumber\\
&&  + a_6 (u_{0,0}+u_{1,0}+u_{0,1} +u_{1,1})
+ a_7 \,=\,0\,,\label{QV}
\end{eqnarray}
where $a_i$ are free complex parameters \cite{Viallet}. By a point fractional-linear
transformation  equation (\ref{QV}) with a generic choice of parameters can be
reduced to Adler's equation, also referred as the Q4 equation in the ABS
classification. Therefore it is not surprising that the Viallet equation (\ref{QV})
has a hierarchy of symmetries \cite{X}. All of the ABS equations can be obtained from
the Viallet equation by a simple specialisation of parameters, it is a convenient way
to represent the ABS equations. The obvious advantage of the Viallet approach is its
applicability to a given equation.

The aim of this paper is to develop a  theory of difference equations, based on the
study of symmetries, conservation laws and (formal) recursion operators. By
integrability of a difference equation we understand the existence of an infinite
hierarchy of its symmetries.  In this paper we consider quadrilateral equations
$Q=0$, where $Q$ is assumed to be an irreducible  affine-linear polynomial over $\C$.
Symmetries, conservation laws, and actually all rational functions which make sense
on solutions of the difference equation are elements of a difference field of
fractions $\cF_Q$ defined  in Section 2.  The affine-linearity of the equation
enables us to introduce the elimination map $\cE$, which reduces any element of
$\cF_Q$ to a rational function of dynamical variables. This map is quite a useful
tool for computations and proofs, since two functions $f$ and $g$ of the lattice
variables are equivalent modulo the difference equation if and only $\cE(f)=\cE(g)$.
In Section 2 we give a self-contained set of basic definitions required in this
paper.

In Section 3 we give a definition of a recursion operator for difference equations.
The recursion operator acting on a seed symmetry of the difference equation generates
an infinite hierarchy of symmetries of that difference equation. We show that the
residues of the powers of the recursion operator are (canonical) densities of
conservation laws for the corresponding difference equation. We can generate an
infinite sequence of canonical conservation laws (both the densities and fluxes).
Similar to the differential case (see \cite{mr86i:58070, mr89g:58092, mr89e:58062}),
these canonical conservation laws can be used as integrability conditions for a
difference equation. They are the necessary conditions for the existence of a
(formal) recursion operator and, ultimately, for an infinite hierarchy of symmetries
of a difference equation.

We present the recursion operators for the Viallet equation and all of the ABS
equations in Section 4. These recursion operators generate infinite hierarchies of
symmetries for every ABS equation. The corresponding canonical conservation laws
provide us with an infinite sequence of densities and corresponding fluxes of
increasing order. The very first term in this sequence was  found by Hydon and Rasin
for all of the ABS equations by a direct computation  \cite{RHcons}. It has been
shown recently that the Gardner method, originally developed for integrable partial
differential equations, can be adapted to difference equations and it was applied for
the production of infinitely many conservation laws for H1 and some other equations,
including  non-symmetrical ones \cite{RasinSchiff,rasin2010}.

Our work was partially inspired by the results obtained by  Levi and Yamilov. In
\cite{LeviYami2009} they started to develop a theory of integrable difference
equations based on the study of their symmetries. The goal of Levi and Yamilov theory
is a classification of difference equations possessing symmetries. They made a good
progress in a formalisation of the theory and they also discussed the issue of dynamical
variables. Levi and Yamilov have found one integrability condition for difference
equations and it does coincide with the first element of our infinite sequence of
canonical conservation laws. Their approach is based solely on symmetries and does
not use the concept of (formal) recursion operator.

We would like to stress that our assumption, that the difference equation is an
affine-linear polynomial and its symmetries are some rational functions, has been
made mainly for the clarity of presentation, and it is suitable for the ABS
equations. This assumption can be removed under some inessential technical
conditions, which guarantee that the difference equation has a unique solution for an
open set of initial data. Obviously all results in this paper are valid for local
analytic functions of the lattice variables.

\section{Basic definitions}
In this section, we introduce the elimination map $\cE$ and define dynamical
variables for quadrilateral affine-linear difference equations. To make the paper
self-contained, we also give the basic definitions of symmetries and conservation
laws of difference equations.
\subsection{Difference equations and dynamical variables}
Difference equations on $\Z^2$ can be seen as a discrete analog of partial
differential equations. Let us denote by $u=u(n,m)$ a complex-valued function
$u:\Z^2\mapsto \C$ where $n$ and $m$ are ``independent variables'' and $u$ will play
the r\^ole of a ``dependent'' variable in a difference equation. Instead of partial
derivatives we have two commuting shift maps $\cS$ and $\cT$ defined as
\[
 \cS: u\mapsto u_{1,0}=u(n+1,m),\qquad \cT: u\mapsto u_{0,1}=u(n,m+1),\qquad \cS^p\cT^q:
 u\mapsto u_{p,q}=u(n+p,m+q).
\]
For uniformity of the notation, it is convenient to denote the ``unshifted'' function
$u$ as $u_{0,0}$.  In the theory of difference equations we shall treat $u_{p,q}$ as variables.
The set of all shifts of the variable $u$ will be denoted by  $U=\{u_{p,q}\,|\, (p,q)\in\Z^2\}$.
For a function $f=f(u_{p_1,q_1},\ldots ,u_{p_k,q_k})$ of  variables $u_{p,q}$
the action of the operators $\cS,\cT$  is defined as
\[
 \cS^i\cT^j (f)=f_{i,j}=f(u_{p_1+i,q_1+j},\ldots ,u_{p_k+i,q_k+j}).
\]

A quadrilateral difference equation can be defined as
 \begin{equation}\label{Qequation}
Q(u_{0,0},u_{1,0},u_{0,1},u_{1,1})=0\ ,
\end{equation}
where $Q(u_{0,0},u_{1,0},u_{0,1},u_{1,1})$ is an irreducible polynomial of the
``dependent variable'' $u=u_{0,0}$ and its shifts. Irreducibility means that 
 $Q$ cannot be factorised and presented as a product of two polynomials. It is
assumed that equation (\ref{Qequation}) is valid at every point $(n,m)\in\Z^2$ and
thus (\ref{Qequation}) represents the infinite set of equations
\begin{equation}\label{Qequation1}
 Q_{p,q}=Q(u_{p,q},u_{p+1,q},u_{p,q+1},u_{p+1,q+1})=0, \qquad (p,\ q)\in\Z^2\ .
\end{equation}

In this paper we shall consider only polynomials $Q$. This restriction is
not very essential for our construction, but allows us to make the presentation more
transparent. Moreover, for quadrilateral  equations we shall assume that $Q$
 is an irreducible
affine-linear polynomial which depends non-trivially on all variables, i.e.
\[ \frac{\partial Q}{\partial u_{i,j}}\ne 0,\ \ \frac{\partial^2 Q}{\partial^2 u_{i,j}}=0,\
\qquad i,j\in\{0,1\},\ Q\in\C[u_{0,0},u_{1,0},u_{0,1},u_{1,1}].
\]
This is true for all of the ABS equations (except the trivial case of the H1 equation
when $\alpha=\beta$).

Let $\C[U]$  be the ring of polynomials of the variables
$U$, where it is assumed that every polynomial has a
finite number of terms. Maps $\cS$ and
$\cT$ are automorphisms of $\C[U]$ and thus $\C[U]$ is a difference ring. We denote
$J_Q=\langle \{Q_{p,q}\}\rangle$ the ideal generated by the difference equation  and
all its shifts (\ref{Qequation1}). Since  $Q$ is an irreducible affine linear polynomial,
the ideal $J_Q$ is prime and radical. For any
$a\in J_Q$ we have $\cS (a)\in J_Q$ and $\cT(a)\in J_Q$. Thus $J_Q$ is a difference
ideal and $\C[U]/J_Q$ is a difference quotient ring.

The fields of rational functions of variables
\[
U_\bs=\{u_{n,0}\,|\, n\in\Z\}, \quad U_\bt=\{u_{0,n}\,|\, n\in\Z\},\quad  U_0=U_\bs\cup U_\bt.
\]
 are denoted as
\[
 \cF_\bs=\C(U_\bs),\qquad \cF_\bt=\C(U_\bt),\qquad  \cF_0=\C(U_0)
\]
respectively. Rational functions of variables $u_{p,q}$ that can be restricted to the
solutions of the difference equation (i.e. the denominators do not vanish on the
solutions) form a field
\[ \cF_Q=\{[a]/[b]\,|\, a,b\in\C[U],\ b\not\in J_Q\}\,, \]
where $[a]$ denotes the class of equivalent polynomials  (two polynomials $f,g\in\C[U]$
are equivalent, denoted by $f\equiv g$, if $f-g\in J_Q$). For $a,b,c,d\in\C[U],\  b,d\not \in J_Q$, the two rational functions
$a/b$ and $c/d$ represent the same element of $\cF_Q$ if $ad-bc\in J_Q$. Such rational functions we shall call
equivalent and denote $a/b\equiv c/d$. The values of equivalent rational functions
coincide for all solutions of the difference equation.

In general, it is quite a tedious computational problem to work with polynomials and rational
functions modulo an ideal. In our case the ideal $J_Q$ is generated by  affine-linear
polynomials $Q_{p,q}$ and we  can uniquely resolve equation $Q=0$ with respect to each
variable
\begin{equation}
 \label{subsQ}
\begin{array}{ll}
u_{0,0}=F(u_{1,0},u_{0,1},u_{1,1}),\qquad &u_{1,0}=G(u_{0,0},u_{0,1},u_{1,1}), \\
u_{0,1}=H(u_{0,0},u_{1,0},u_{1,1}),\qquad &u_{1,1}=M(u_{0,0},u_{1,0},u_{0,1}).
\end{array}
\end{equation}
Since  $Q$ is an affine linear polynomial, functions  $F,G,H$
and $M$ are rational functions of their arguments.
Equations (\ref{subsQ}) enable us  recursively and uniquely to
express any variable $u_{p,q}$ in terms of the variables $U_0=U_\bs\cup U_\bt$. 
\begin{Def}\label{Def1}
 For elements of $U$ the {\em elimination map} $\cE:U\mapsto\C(U_0)$ is defined recursively:
\begin{equation}\label{Emap}
\begin{array}{ll}
\forall p\in\Z,\ \qquad &\cE(u_{0,p})=u_{0,p},\qquad  \cE(u_{p,0})=u_{p,0}\, ,\\
{\rm if}\ p>0,q>0,\  \quad &\cE(u_{p,q})=M(\cE(u_{p-1,q-1}),\cE(u_{p,q-1}),\cE(u_{p-1,q}))\,,\\
{\rm if}\ p<0,q>0,\  \quad &\cE(u_{p,q})=H(\cE(u_{p,q-1}),\cE(u_{p+1,q-1}),\cE(u_{p+1,q}))\,,\\
{\rm if}\ p>0,q<0,\  \quad &\cE(u_{p,q})=G(\cE(u_{p-1,q}),\cE(u_{p-1,q+1}),\cE(u_{p,q+1}))\,,\\
 {\rm if}\ p<0,q<0,\ \quad &\cE(u_{p,q})=F(\cE(u_{p+1,q}),\cE(u_{p,q+1}),\cE(u_{p+1,q+1}))\, .
 \end{array}
\end{equation}
For  polynomials $f(u_{p_1,q_1},\ldots ,u_{p_k,q_k})\in\C[U]$ the   elimination map $\cE:\C[U] \mapsto \C(U_0)$ is defined as 
$$\cE:f(u_{p_1,q_1},\ldots ,u_{p_k,q_k})\mapsto f(\cE(u_{p_1,q_1}),\ldots ,\cE(u_{p_k,q_k}))\in\C(U_0).$$ 

\noindent
For rational functions $a/b,\
a,b\in\C[U], \ b\not\in J_Q$ the elimination map $\cE$ is defined as 
$$\cE:a/b \mapsto\cE(a)/\cE(b).$$
\end{Def}

Variables $U_0$ we shall call the {\em dynamical} variables. it is easy to see that
$\cE(u_{p,q})$ is a rational function of $|p|+|q|+1$  dynamical variables
\begin{equation}
\cE(u_{p,q})\in\C(\{u_{n,0},u_{0,m}\,|\, 0\le|n-p|\le |p|,\ 0\le|m-q|\le |q|\}).
\end{equation}
The choice of the set of dynamical variables is not unique. Another popular choice is
the so-called ``staircase'' set $\{u_{n,n},u_{n,n+1}\,|\, n\in\Z\}$ which will not be
used in this paper.
 
The  elimination map $\cE:\C[U] \mapsto \C(U_0)$ is a difference ring homomorphism,
its kernel is $J_Q$, and its image is a difference ring isomorphic to the factor ring
$\C[U]/J_Q$. The field $\C(U_0)$ is a difference field with automorphisms
$\cE\circ\cS$ and $\cE\circ\cT$. The elimination map $\cE:\cF_Q \mapsto \C(U_0)$ is a
difference field isomorphism. It is a useful tool to establish whether two  rational 
functions $f,g$ of variables $U$ are equivalent  (i.e. represent the same element of $\cF_Q$):
$f\equiv g$  if and only if $\cE(f)=\cE(g)$.
 Later on we shall  use
logarithms  of rational functions. The action of $\cE$ on $\ln (a), \ a\in\cF_Q,\
a\not\in J_Q$ is defined as $\cE(\ln (a))=\ln(\cE(a))$
and thus we shall say that $\ln (a) \equiv\ln(b)$ if and only if $\ln(\cE(a))=\ln(\cE(b))$.

 To illustrate the above construction, let us consider the difference equation $Q=0$ with
\begin{equation}\label{H1equationS}
 Q=(u_{0,0}-u_{1,1})(u_{1,0}-u_{0,1})-\alpha +\beta,\qquad \alpha,\beta\in\C,
\end{equation}
which is known as the H1 equation in the ABS classification or the discrete potential
KdV equation. First we should assume that $\alpha\ne\beta$, otherwise the
polynomial $Q$ is reducible. It follows from Definition \ref{Def1} that
\begin{eqnarray*}
 \cE(u_{1,1})&=&M(u_{0,0},u_{1,0},u_{0,1})\ =\ u_{0,0}-\frac{\alpha-\beta}{u_{1,0}-u_{0,1}},\\
 \cE(u_{1,2})&=&\frac {u_{0,1}^2 u_{0,2}-u_{0,0} u_{0,1}^2-(\alpha-\beta)
u_{1,0}+u_{0,0} u_{0,1} u_{1,0}-u_{0,1} u_{0,2} u_{1,0}} {\beta-\alpha -u_{0,0}
u_{0,1}+u_{0,1} u_{0,2}+u_{0,0} u_{1,0}-u_{0,2} u_{1,0}}, \ \ldots\ .
\end{eqnarray*}
Two rational functions
\[
f=\frac{u_{0,0}-u_{1,1}}{u_{1,0}-u_{-1,0}}\, ,\quad g=\frac{\alpha-\beta}
{(u_{1,0}-u_{-1,0})(u_{1,0}-u_{0,1})}
\]
are equivalent $f\equiv g$ since
\[
 \frac{u_{0,0}-u_{1,1}}{u_{1,0}-u_{-1,0}}-\frac{\alpha-\beta}{(u_{1,0}-u_{-1,0})(u_{1,0}-u_{0,1})}=
\frac{Q}{(u_{1,0}-u_{-1,0})(u_{1,0}-u_{0,1})}
\]
and it is easy to check that $\cE(f)=\cE(g)=g$.

\subsection{Symmetries and conservation laws of difference equations}
\begin{Def} \label{DefSymmetry}
Let $Q=0$ be a difference equation. Then $K\in\cF_Q$ is called a {\em generator of an
infinitesimal symmetry} (or simply, a {\em symmetry}) of the difference equation if
\begin{equation}\label{sym_def}
D_Q(K)\equiv 0.
\end{equation}
Here $D_Q$ is the Frech\'et derivative of $Q$ defined as
\begin{equation}\label{frechet}
D_Q=\sum_{i,j}Q_{u_{i,j}} \cS^i \cT^j\ ,\qquad Q_{u_{i,j}}=\frac{\partial Q}{\partial u_{i,j}}.
\end{equation}
\end{Def}
The sum in (\ref{frechet}) is finite for a given $Q$. For a quadrilateral equation it
has only four terms. Here $D_Q(K)$ is equal to zero as element of $\cF_Q$, i.e.
$D_Q(K)\subset J_Q$. The way to check it is to apply the elimination map: thus, $K$
is a symmetry if $\cE(D_Q(K))=0$.

If $K$ is a symmetry and  $u=u(n,m)$ is a  solution of a difference equation
(\ref{Qequation}), then  there is an infinitesimal transformation of solution $u$:
\[ \hat{u}=u+\epsilon K\]
satisfying
\[Q(\hat{u}_{0,0},\hat{u}_{1,0},\hat{u}_{0,1},\hat{u}_{1,1})\equiv{\cal O}(\epsilon^2).\]

If the difference equation $Q=0$ admits symmetries, then they form a Lie algebra.
Indeed, with a symmetry $K\in \cF_Q$ we can associate an evolutionary derivation of
$\cF_Q$ (or a vector field on $\cF_Q$):
\[ X_K=\sum_{(p,q)\in\Z^2}K_{p,q}\frac{\partial }{\partial u_{p,q}},
\]
where $K_{p,q}=\cS^p\cT^q(K)$. A derivation (a vector field)  $X_K$ is called
evolutionary if it commutes with the shift automorphisms:  $\cS X_K=X_K\cS, \ \cT
X_K=X_K\cT$. It follows from (\ref{sym_def}) that for any $a\in J_Q$ we have
$X_K(a)\in J_Q$ and thus the evolutionary derivation $X_K$ is defined correctly on
the difference field of fractions $\cF_Q$.

If the difference equation admits  two symmetries $F$ and $G$, then the  commutator
of the corresponding evolutionary derivations is an evolutionary derivation of
$\cF_Q$
\[ X_F X_G-X_G X_F=X_H ,\]
where $H=[F,G]$, with  $[F,G]$ denoting the {\em Lie bracket}
\begin{equation}\label{liebracket}
 [F,G]=X_F(G)-X_G(F)=D_G(F)-D_F(G)\in\cF_Q.
\end{equation}
It is easy to see that $H$ is a symmetry, indeed
\[ D_Q(H)=X_H(Q)=X_F X_G(Q)-X_G X_F(Q)=X_F D_Q(G)-X_G D_Q(F)\equiv 0.\]
Symmetries form a linear space over $\C$. The bracket (\ref{liebracket}) is linear
with respect to each argument, it is skew-symmetric and satisfies the Jacobi
identity. The Lie algebra of symmetries of the difference equation $Q=0$ will be
denoted as $\gA_Q$. Existence of an infinite dimensional Lie algebra $\gA_Q$  is a
characteristic property of integrable equations and can be taken as a definition of
integrability.

We would like to emphasise that a symmetry $K$ is an element of $\cF_Q$, so it  can
be represented by a function $\cE(K)\in \cF_0$ of variables $U_0$. As an element of
$\cF_0$, a symmetry $K$ depends on a finite set of variables $\{u_{n,0}\,|\, N_1 \le
n \le N_2 \}$ and $\{u_{0,m}\,|\, M_1\le n \le M_2\}$, and thus can be characterised
by four integers $(N_1,N_2,M_1,M_2)$. It is straightforward to check that, if
$$K=K(u_{N_1,0},\ldots,u_{N_2,0},u_{0,M_1},\ldots,u_{0,M_2})$$ is a symmetry of a
quadrilateral equation, then it is a sum of two functions
$$K=K_{\bs}(u_{N_1,0},\ldots,u_{N_2,0})+K_{\bt}(u_{0,M_1},\ldots,u_{0,M_2})\ .$$ In
particular, for the ABS equations  it is known \cite{X,TTX} that, any
``five point symmetry'' $K=K(u_{0,0}, u_{-1,0},u_{0,-1}, u_{1,0}, u_{0,1})$ is a sum
of two symmetries $K=K_{\bs}( u_{-1,0}, u_{0,0}, u_{1,0})+K_{\bt}( u_{0,-1},
u_{0,0},u_{0,1})$. Thus, the problem of the symmetry description is likely to be reduced
to the study of the symmetries depending on ``one-dimensional'' set of variables $U_\bs$
or $U_\bt$. Moreover, all ABS equations and the Viallet equation possess infinite
hierarchy of such symmetries.

Let $f=f(u_{N_1,0},\ldots , u_{N_2,0})\in\cF_\bs$ and
$
\partial f/\partial u_{N_1,0}\ne 0 ,\ \partial f/\partial u_{N_2,0}\ne 0$, then
the order of  $f$ is defined as  $\mbox{ord}_\bs (f)=(N_1,N_2)$. For example,
\begin{equation}
 \label{SigmaH1}
K^{(1)}=(u_{1,0}-u_{-1,0})^{-1}
\end{equation}
is the generator of a symmetry for equation (\ref{H1equationS}) and
$\mbox{ord}_\bs(K^{(1)})=(-1,1)$. Similarly, the order of  $g=g(u_{0,M_1},\ldots ,
u_{0,M_2})\in\cF_\bt$ is defined as $\mbox{ord}_\bt (g)=(M_1,M_2)$.

\begin{Def}
\begin{enumerate}\item
A pair of functions  $\rho,\sigma\in\cF_Q$ is called a {\em conservation law}
for the difference equation (\ref{Qequation}), if
\begin{equation}
 \label{conslaw}
(\cT-{\bf 1})(\rho)\equiv (\cS-{\bf 1})(\sigma).
\end{equation}
Functions $\rho$  and $\sigma$ will be referred to as the  density and the flux
of the conservation law and ${\bf 1}$ denotes the identity map.
\item A conservation law is called {\em trivial}, if functions $\rho$ and $\sigma$
are components of
a (difference) gradient of some element $H\in \cF_Q$, i.e.
\[ \rho=(\cS-{\bf 1})(H),\qquad \sigma=(\cT-{\bf 1})(H). \]
\item A non-constant element $C\in\cF_Q$ is called $\bs$--constant,
(respectively $\bt$--constant)
if $\cS(C)=C$ (respectively $\cT(C)=C$). Elements of $\cF_Q$ which are invariant with
respect to the both shifts are called constants of $\cF_Q$.
\end{enumerate}
\end{Def}

Existence of non-trivial conservation laws and/or $\bs$ or $\bt$--constants for a
difference equation is a rather exceptional property. The ABS equations (and many
other integrable difference equations)
 possess infinitely many nontrivial conservation laws. Similar to the
differential case \cite{mr1845643}, the  existence of $\bs$ or $\bt$--constants
suggests that the difference equation is  linearizable. For instance the linear
difference equation with $Q=u_{0,0}-u_{1,0}-u_{0,1}+u_{1,1}$ has a $\bt$--constant
$f(u_{1,0}-u_{0,0})$ and a $\bs$--constant $g(u_{0,1}-u_{0,0})$, where $f,g$ are
arbitrary functions.

\begin{Def} Let
$\im(\cS-{\bf 1})$ and $\im(\cT-{\bf 1})$  denote the images of the maps $\cS-{\bf
1}:\cF_Q\mapsto \cF_Q$ and $\cT-{\bf
1}:\cF_Q\mapsto \cF_Q$ respectively.
\begin{enumerate}
\item Two elements $\rho,\varrho\in\cF_Q$ are $\bs$-equivalent ($\rho\cong_\bs {\varrho}$)
if $\rho-\varrho\in
\im(\cS-{\bf 1})$. Elements of the factor space $\cF_Q/\im(\cS-{\bf 1})$ we
shall call $\bs$-functionals, or densities.
\item Two elements $\sigma,\varsigma\in\cF_Q$ are {\bt}-equivalent ($\sigma\cong_\bt \varsigma$)
if $\sigma-\varsigma\in \im(\cT-{\bf 1})$. Elements of the factor space
$\cF_Q/\im(\cT-{\bf 1})$ we shall call $\bt$-functionals, or fluxes.
\end{enumerate}
\end{Def}

In general the problem to establish whether two elements of $\cF_Q$ are $\bs$- or
$\bt$-equivalent is difficult. This problem can be easily solved for densities
belonging to $\cF_\bs$ (or fluxes belonging to $\cF_\bt$). Moreover, in the next
Section we will show that an integrable quadrilateral equation (such as the the
Viallet equation) possesses an infinite hierarchies of canonical conservation laws
such that their densities  are elements of $\cF_\bs$.

There is a criterion to determine whether two elements of $\cF_\bs$ are equivalent or
not. This is based on the notion of variational derivative (Euler's operator).

\begin{Def}
 Let $f\in\cF_\bs$ has order $(N_1,N_2)$, then the variational derivative
 $\delta_\bs$ of $f$ is defined as
\begin{equation}
 \label{varder}
\delta_\bs(f)=\sum_{k=N_1}^{N_2} \cS^{-k}\left(\frac{\partial f}{\partial u_{k,0}}\right).
\end{equation}
\end{Def}
Two elements $ \rho ,\varrho\in\cF_\bs$ are equivalent $\rho \cong_\bs\varrho$ if and
only if $\delta_\bs (\rho)=\delta_\bs (\varrho)$. In particularly, if density
$\rho\in\cF_\bs$ is trivial, then $\delta_\bs (\rho)=0$. The order of a density
$\rho\in\cF_\bs$ is defined as $\mbox{ord}_{\delta_\bs}( \rho)=N_2-N_1$, where
$(N_1,N_2)=\mbox{ord}(\delta_\bs(\rho))$. Equivalent densities have the same order.
For example the densities of conservation laws for the H1 equation
(\ref{H1equationS})
\[
\rho_0=\ln (u_{1,0}-u_{-1,0})\quad \mbox{and}\quad
\rho_1=\frac{1}{(u_{1,0}-u_{-1,0})(u_{2,0}-u_{0,0})},
\]
are of  orders $4$ and $6$ respectively.

\section{Recursion operator and canonical conservation laws}
In this section we give  definitions of ($\bs$- and $\bt$- ) difference and
pseudo-difference operators. We adapt the concept of recursion operators to
difference equations and give the criteria for a pseudo-difference operator to be a
recursion operator of a quadrilateral difference equation. Using Laurent formal
series and their residues we derive infinitely many integrability conditions for
quadrilateral difference equations.

\subsection{Recursion operators of difference equations}\label{sec24}

In order to proceed we need to define $\bs$- and $\bt$-pseudo-difference operators.

\begin{Def} (1) A $\bs$-difference operator $B$ of order $N$ with coefficients in
$\cF_Q$ is a finite sum of the form
\begin{equation}\label{operB}
B= b_N \cS^{N}+b_{N-1} \cS^{N-1}+\cdots +b_M \cS^{M},\qquad b_k\in\cF_Q, \ \ M\le N,\
\ N,M\in\Z.
\end{equation}
(2) An operator $L$ is called a $\bs$-pseudo-difference operator if it is either a finite
sum of terms
\begin{equation}\label{psudo_diff}
 L=\sum B_n\circ  C^{-1}_n
\end{equation}
(where $B_n$ and $C_n$ are $\bs$-difference operators with coefficients in
$\cF_Q$ and $\circ$ denotes their composition) or a finite composition of
$\bs$-pseudo-difference operators of the form (\ref{psudo_diff}).
\end{Def}

In the above definition we used the shift map $\cS$, and thus we have defined the
$\bs$-difference, $\bs$-pseudo-difference operators. Using the shift map $\cT$
instead, one can  define the $\bt$-difference and pseudo-difference operators. We
shall omit $\bs$- (or $\bt$-) and simply say difference or pseudo-difference
operators if it is clear from the context.

There is a natural action of difference operators on all elements of $\cF_Q$ but the action
of pseudo-difference operators is not generally defined. For a pseudo-difference
operator $L$ (\ref{psudo_diff}) it can be defined on those elements of $a_i\in\cF_Q$
that belong to the intersection of the image spaces of the difference operators
$C_n$, so that the action of operators $C_n^{-1}(a_i)$ is defined, i.e.  $C_n^{-1}(a_i)\in\cF_Q$.
A linear space that is a $\C$-span of all such elements $a_i$ is a domain of the
pseudo-difference operator denoted by ${\rm Dom}(L)$.

By a recursion operator $\gR$ of a difference equation $Q=0$ we shall understand a
$\bs$-pseudo-difference operator  such that
\[
\gR:{\rm Dom}(\gR)\cap \gA_Q\mapsto \gA_Q,
\]
where $\gA_Q$ is the linear space of symmetries of this difference equation. In other
words, if the action of $\gR$ on a symmetry $K\in\cF_0$ is defined, i.e. $\gR (K)\in
\cF_Q$, then $\gR (K)$ is a symmetry of the same difference equation.  Similarly,  we
define a  $\bt$-pseudo-difference recursion  operator $\hat{\gR}:{\rm
Dom}(\hat{\gR})\cap  \gA_Q \mapsto  \gA_Q$.  The operator of multiplication by a
constant  is a {\em trivial} recursion operator. In what
follows we shall assume that $\gR,\hat{\gR}$ are nontrivial.

\begin{The}\label{The2} Let $Q(u_{0,0},u_{1,0},u_{0,1},u_{1,1})=0$ be a difference equation.

{\rm (i)} If there exist two  $\bs$--pseudo-differential operators $\gR$ and $\gP$ such that
\begin{equation}\label{rec}
 D_{Q} \circ \gR=\gP \circ D_{Q},
\end{equation}
then $\gR$ is a recursion operator of the difference equation.

{\rm (ii)} Relation (\ref{rec}) is valid if and only if
\begin{equation}\label{recursion_eq}
 \cT(\gR)-\gR=[\Phi\circ \gR,\Phi^{-1}]\ ,
\end{equation}
where $\Phi=(Q_{u_{1,1}}\cS+Q_{u_{0,1}})^{-1}\circ( Q_{u_{1,0}}\cS+Q_{u_{0,0}})$, and
the operator $\gP$ satisfies
\begin{equation}\label{gP1}
 \gP=(Q_{u_{1,0}}\cS+Q_{u_{0,0}})\circ \gR\circ (Q_{u_{1,0}}\cS+Q_{u_{0,0}})^{-1}.
\end{equation}
\end{The}
\noindent
{\bf Proof}. (i) If $K$ is a symmetry and $\gR (K)\in\cF_Q$ it follows from (\ref{rec}) that
$D_Q\gR(K)=\gP D_Q(K)\equiv 0$ and thus $\gR (K)$ is a symmetry.

To prove (ii), we represent $D_Q$ in the factorised form
\[ D_Q=Q_{u_{1,1}}\cS\circ \cT+Q_{u_{0,1}}\cT+Q_{u_{1,0}}\cS+Q_{u_{0,0}}=(Q_{u_{1,1}}
\cS+Q_{u_{0,1}})\circ(\cT+\Phi)\, ,\]
where
\begin{equation}
 \Phi=(Q_{u_{1,1}}\cS+Q_{u_{0,1}})^{-1}\circ( Q_{u_{1,0}}\cS+Q_{u_{0,0}}).\label{Phi}
\end{equation}
Then (\ref{rec}) can be rewritten as
\begin{equation}\label{eqR1}
\cT(\gR)\circ\cT+\Phi\circ\gR= (Q_{u_{1,1}}\cS+Q_{u_{0,1}})^{-1}\circ \gP\circ
(Q_{u_{1,1}}\cS+Q_{u_{0,1}})\circ (\cT+\Phi)\ ,
\end{equation}
which is equivalent to
\begin{equation}\label{eqR100}
 \cT(\gR)=(Q_{u_{1,1}}\cS+Q_{u_{0,1}})^{-1}\circ \gP\circ (Q_{u_{1,1}}\cS+Q_{u_{0,1}})
\end{equation}
and
\begin{equation}\label{ttR}
 \Phi\circ\gR=(Q_{u_{1,1}}\cS+Q_{u_{0,1}})^{-1}\circ
 \gP\circ (Q_{u_{1,1}}\cS+Q_{u_{0,1}})\circ \Phi.
\end{equation}
Using (\ref{Phi}) we can rewrite (\ref{ttR}) in the form (\ref{gP1}).
It follows from (\ref{eqR100}) and (\ref{ttR}) that
\begin{equation}\label{PhiR}
 \Phi\circ \gR=\cT(\gR)\circ\Phi
\end{equation}
The latter can be rewritten in the form (\ref{recursion_eq}).\hfill $\blacksquare$

In the same way one can prove the similar statement in the $\bt$-direction.
\begin{The}\label{The2T} Let $Q(u_{0,0},u_{1,0},u_{0,1},u_{1,1})=0$ be a difference equation.

{\rm (i)} If there exist two  $\bt$--pseudo-differential
operators $\hat{\gR}$ and $\hat{\gP}$ such that
\begin{equation}\label{recT}
 D_{Q} \circ \hat{\gR}=\hat{\gP} \circ D_{Q},
\end{equation}
then $\hat{\gR}$ is a recursion operator of the difference equation.

{\rm (ii)} Relation (\ref{recT}) is valid if and only if
\begin{equation}\label{recursion_eqT}
 \cS(\hat{\gR})-\hat{\gR}=[\Psi\circ \hat{\gR},\Psi^{-1}]
\end{equation}
where $\Psi=(Q_{u_{1,1}}\cT+Q_{u_{1,0}})^{-1}\circ( Q_{u_{0,1}}\cT+Q_{u_{0,0}})$, and the
operator $\hat{\gP}$ can be written as
\[
 \hat{\gP}=(Q_{u_{0,1}}\cT+Q_{u_{0,0}})\circ \hat{\gR}\circ (Q_{u_{0,1}}\cT
 +Q_{u_{0,0}})^{-1}.
\]
\end{The}
As in the case of partial differential equations, we show that if $\gR$ is a
recursion operator, so is $\gR^n$ for all $ n\in\Z$ in the following statement.
\begin{Cor}
1.  Under the conditions of {\rm (i)} in Theorem \ref{The2}, the pseudo-difference
operator $\gR$ satisfies the following equations
\begin{equation}\label{Rn_eq}
\cT(\gR^n)-\gR^n=[\Phi\circ \gR^n,\Phi^{-1}],\qquad  n\in\Z,
\end{equation}

\noindent
2. Under the conditions of {\rm (i)} in  Theorem \ref{The2T},
 the pseudo-difference operator $\hat{\gR}$ satisfies equations
  \begin{equation}\label{Rn_eqT}
 \cS(\hat{\gR}^n)-\hat{\gR}^n=[\Psi\circ \hat{\gR}^n,\Psi^{-1}]\qquad  n\in\Z,
\end{equation}
\end{Cor}
{\bf Proof}. For the first part, it follows from (\ref{rec}) that $ D_{Q} \circ
\gR^n=\gP^n \circ D_{Q}, n\in\Z$. Thus, we can apply Theorem \ref{The2} to $\gR^n$
and $\gP^n$ to produce (\ref{Rn_eq}) from (\ref{recursion_eq}). The proof of the
second part of the Corollary is similar. \hfill$\blacksquare$

In all our definitions and statements there is an obvious symmetry between ``$\cT$''
and ``$\cS$'' objects (such as symmetries, pseudo-differential operators, symplectic
and  Hamiltonian operators). In what follows we shall concentrate on ``$\cS$''
objects, but formulate some statements for both if required.

\subsection{Canonical conservation laws and integrability conditions} \label{Rec&CL}

\begin{Def}
 A  formal (Laurent\footnote{One can also define a formal  Taylor series of order
 $-N$ as a semi-infinite sum
\[
 C=c_{-N}\cS^{-N}+ c_{1-N}\cS^{1-N}+\cdots +c_{-1}\cS^{-1}+c_0+c_{1}\cS+\cdots \, ,
 \qquad c_N\ne 0,\ \ c_k\in\cF_Q,\ \ N\in\Z
\]
to develop a similar theory based on the Taylor series (or both). But in the case when the
recursion operator is a composition of a Hamiltonian and symplectic operators it can
be shown that the both approaches give equivalent results. Thus in applications
to the ABS and Viallet equations we restrict ourselves to Laurent formal series only. Study of the both (Laurent and Taylor) series make sense for asymmetric difference equations.
\label{footnote1}}) series of order $N$ is defined as a formal semi-infinite sum
\begin{equation}\label{formser}
A=a_{N}\cS^N+ a_{N-1}\cS^{N-1}+\cdots +a_1\cS+a_0+a_{-1}\cS^{-1}+\cdots \, ,
\qquad a_N\ne 0,\ \ a_k\in\cF_Q,\ \ N\in\Z.
\end{equation}
\end{Def}

Laurent  formal series form a skew-field.  Sums and products (compositions) of formal
series are formal series. The product is associative, but not commutative. For any
formal series $A$ there exists a formal series $A^{-1}$ such that $A\circ A^{-1}=
A^{-1}\circ A={\bf 1}$. In order to find the first $n$ coefficients of $A^{-1}$ one
needs to know exactly the first $n$ coefficients of $A$. The same holds for the
composition of two formal series. For instance, for two Laurent formal series $A$
defined by (\ref{formser}), and $B=b_M \cS^M+b_{M-1}\cS^{M-1}\ldots $, we have
\[ A\circ B=\sum_{k=0}^\infty c_{N+M-k}\cS^{N+M-k},\qquad c_k=
\sum_{q=0}^{N+M-k}a_{N-q}\cS^{N-q}(b_{q+k-N})\,  \in \cF_Q.
\]

Any pseudo-difference operator $B$ can be uniquely represented by a Laurent formal
series $B_L$. For example for $B=(a\cS+b)^{-1}$ we have
\[
 B_L=\alpha_{-1}\cS^{-1}+\alpha_{-2}\cS^{-2}+\alpha_{-3}\cS^{-3}+\cdots,
\]
where the coefficients $\alpha_k\in\cF_Q$ can be found recursively:
\[
 \alpha_{-1}=\cS^{-1}\left(\frac{1}{a}\right),\  \alpha_{-n}=-\cS^{-1}
 \left(\frac{\alpha_{1-n}b}{a}\right).
\]

\begin{Def} Let $A_L$  denote the Laurent series representations of a pseudo-differential
operator $A$. Then ${\rm ord}A_L$ is  called the Laurent  order of $A$.
\end{Def}

Now we introduce the residue of a formal series, as well as a discrete version of
Adler's theorem \cite{mr80i:58026}, which will be used consequently to produce
integrability conditions for quadrilateral difference equations.

\begin{Def}
Let $A$ defined by (\ref{formser}) be a formal series of order $N$.
The residue ${\rm res}(A)$ and the logarithmic residue ${\rm res}\ln(A)$ are defined
as
\[ {\rm res}(A)=a_0,\qquad {\rm res}\ln(A)=\ln (a_N)\, .\]
\end{Def}
We shall need the following difference analog of the Adler Theorem \cite{mr80i:58026}:
\begin{The}\label{Adler}
Let $A=a_{N} \cS^N+a_{N-1}\cS^{N-1}\cdots $ and $B=b_{M} \cS^M+b_{M-1}\cS^{M-1}\cdots
$ be two Laurent formal series of order $N$ and $M$ respectively. Then
 \[
 {\rm res}[A,B]=(\cS-{\bf 1})(\sigma(A,B)),
\]
where
\[
\sigma (A,B)=\sum_{n=1}^N \sum_{k=1}^n \cS^{-k}(a_{-n})\cS^{n-k}(b_n)-\sum_{n=1}^M
\sum_{k=1}^n \cS^{-k}(b_{-n})\cS^{n-k}(a_n)\in\cF_Q.
\]
\end{The}
{\bf Proof}. Only commutators of  the form
$[a_{p}\cS^{p},b_{-p}\cS^{-p}], \ p\in\Z$ contribute to the residue. Assuming $p>0$,
we get
\[
 [a_{p}\cS^p,b_{-p}\cS^{-p}]=a_{p}\cS^p(b_{-p})-\cS^{-p}(a_{p})b_{-p}
 =(\cS^p-{\bf 1})(\cS^{-p}(a_{p})b_{-p})=
(\cS-{\bf 1})\sum_{k=1}^p \cS^{p-k}(\cS^{-p}(a_{p})b_{-p}).\hfill \blacksquare
\]

For first order formal series $A$   the first $n$
coefficients of the series completely define the  first $n$ residues $\rho_0={\rm res}\ln A,
\rho_1={\rm res}A,\ldots ,\rho^{n-1}={\rm res}A^{n-1}$.

In the theory of pseudo-differential formal series (see for example
\cite{mr86i:58070, mr89e:58062, mr93b:58070}) for a series
$A=aD_x^N+bD_x^{N-1}+\cdots$ of order $N$ it is always possible to find the $N$-th
root, i.e. a formal series
$B=A^{\frac{1}{N}}=a^{\frac{1}{N}}D_x+\left(\frac{1}{N}a^{\frac{1}{N}-1}b-
\frac{N-1}{N}a^{\frac{1}{N}-1}D_x(a)\right)+\cdots $ such that  $B^N=A$ and,
consequently, to parameterize the coefficients of $A$ by a sequence of canonical
residues ${\rm res}\ln B, {\rm res}B,  {\rm res}B^2,\ldots $. In the case of formal
difference series in general it is not possible to find the $N$-th root of a series $A=a_N\cS^N+a_{N-1}\cS^{N-1}+\cdots$, i.e. a first order series 
$B=\alpha_1\cS+\alpha_0+\cdots$ with the coefficients in $\cF_Q$, such that $B^N=A$. Indeed, 
computations show that to
determine the coefficient $\alpha_k$ one needs to solve  equations of the form
\[
 \prod_{i=0}^{N-1}\cS^i(\alpha_1)=a_N,\qquad    \alpha_k+\cS(\alpha_k)+\cdots +\cS^{N-1}(\alpha_k)=g_k,\quad k=0,-1,-2,\ldots
\]
for some $g_k\in\cF_Q$. Obviously, a solution $\alpha_k\in\cF_Q$ does not exist if
$g_k\not\in {\rm Im}({\bf 1}+\cS+\cdots \cS^{N-1})$.

\begin{The}\label{The4} If a difference equation  possesses a  recursion
operator $\gR,\ {\rm ord}_L(\gR)=N>0$, then it
has infinitely many {\rm canonical} conservation laws
\[ (\cT-{\bf 1})\rho_{nN}=(\cS-{\bf 1})\sigma_{nN}, \quad n=0,1,2,\ldots\]
 with {\rm canonical} conserved densities
\begin{equation}\label{rho_n}
 \rho_0={\rm res}\ln \gR_L,\quad \rho_{nN}={\rm res}\gR^n_L,\ n>0.
\end{equation}
\end{The}
{\bf Proof}. Equation (\ref{Rn_eq}) is obviously satisfied if we replace $\Phi$ and
$\gR$ by the corresponding Laurent formal series $\Phi_L$ and $\gR_L$. For $n\ge 1$,
we compute the residue of equation  (\ref{Rn_eq})  to arrive at
\begin{equation}\label{Rnconslaws}
\cT({\rm res}\gR^n_L)-{\rm res}\gR^n_L={\rm res}([\Phi_L\circ \gR^n_L,\ \ \Phi^{-1}_L])
=(\cS-{\bf 1})\sigma(\Phi_L\circ \gR^n_L,\Phi^{-1}_L).
\end{equation}
It follows from Theorem \ref{Adler} that
\begin{equation}\label{sigma_n}
 \sigma_{nN}=\sigma(\Phi_L\circ \gR^n_L,\Phi^{-1}_L),\qquad \sigma_n\in\cF_Q.
\end{equation}
For $n=0$ , we rewrite (\ref{PhiR}) as
\begin{equation*}
\Phi\circ\gR\circ\Phi^{-1}\circ\gR^{-1}=\cT(\gR)\circ\gR^{-1}\,.
\end{equation*}
and compute its logarithmic residue. This leads to
\begin{equation}
 \label{sigma0}
\cT({\rm res}\ln \gR_L)-{\rm res}\ln\gR_L=(\cS-{\bf 1})
\sum_{k=0}^{N-1}\cS^{k-1}(\ln\alpha_0 ),
\end{equation}
where $\alpha_0$ is the first coefficient in the expansion
$\Phi_L=\alpha_0+\alpha_1\cS^{-1}+\cdots$, (see (\ref{Phi_alphabeta})). \hfill
$\blacksquare$

In the above Theorem, densities and fluxes are counted modulo $N$, where $N$ is the
Laurent order of the recursion operator.

If a recursion operator is known, then  Theorem \ref{The4} gives us a completely
algorithmic way to find explicitly a sequence of conservation laws, including both
the densities $\rho_k$ and the corresponding fluxes $\sigma_k$. The residues of
powers  a formal series are easy to compute. For instance, if
\[ \gR=r_1\cS+r_0+r_{-1}\cS^{-1}+r_{-2}\cS^{-2}+r_{-3}\cS^{-3}+
\]
then
\begin{eqnarray}
{\rm res}\ln \gR&=&\ln r_1,\quad  {\rm res} \gR=r_0,\quad {\rm res} \gR^2= \cS^{-1}(r_1) r_{-1}
 + r_0^2 + r_{1}\cS^{-1}(r_{-1}), \nonumber\\
{\rm res} \gR^3&=&\cS^{-2}(r_1)\cS^{-1}(r_1) r_{-2}+\cS^{-1}(r_0)\cS^{-1}(r_1)r_{-1}
 + 2\cS^{-1}(r_1)r_{-1}r_0   \label{ResiduesOrder1} \\
&& + \cS^{-1}(r_1)r_1\cS(r_{-2})+r_0^3 + 2r_0 r_1\cS(r_{-2})+ r_1 \cS(r_{-1})\cS(r_0)
+ r_1 \cS(r_1)\cS^2(r_{-2}). \nonumber
\end{eqnarray}

\begin{Pro} \label{PropIntegrCon1}
If a recursion operator $\gR$ is represented by a first order formal series
$\gR_L=r_1 \cS+r_0+r_{-1}\cS^{-1}+\cdots$, then
\begin{eqnarray}\label{cond0}
({\rm i})&\qquad & (\cT-{\bf 1})(\ln r_1)=(\cS-{\bf
1})\cS^{-1}\left(\ln\frac{Q_{u_{1,1}}}{Q_{u_{1,0}}}\right),\\ \label{cond1}
({\rm ii})&\qquad &  (\cT-{\bf 1})( r_0)=(\cS-{\bf 1})\cS^{-1}(r_1 F),\\ \label{cond2}
({\rm iii})&\qquad &(\cT-{\bf 1})(
r_{-1}\cS^{-1}(r_1)+r_0^2+r_1\cS(r_{-1}))=(\cS-{\bf 1})
(\sigma_2),
\end{eqnarray}
where
\[
\sigma_2 \,= \, \cS^{-1}(r_1\, F)\, \left\{ \cS^{-1}(r_0)\, +\, r_0\,
-\,\cS^{-2}\left( r_1 \,F\right)\right\}\, -\, ( 1+\cS^{-1}) \left( r_1 \,G\,
\cS^{-1} \left(r_1 \, F\right) \right)\,,
\]
and $F,G$ denote
\[
 F=\frac{Q_{u_{0,1}}
\cS^{-1}(Q_{u_{1,0}})-
Q_{u_{0,0}}\cS^{-1}(Q_{u_{1,1}})}{Q_{u_{1,0}}\cS^{-1}(Q_{u_{1,1}})},\qquad
G=\frac{Q_{u_{0,0}}}{Q_{u_{1,0}}}.
\]
\end{Pro}
{\bf Proof}.
We expand $\Phi$ and $\Phi^{-1}$ (\ref{Phi}) in the formal Laurent series
\[
\Phi_L=\alpha_0+\alpha_1\cS^{-1}+\alpha_2\cS^{-2}+\ldots\,,\quad \Phi^{-1}_L
=\beta_0+\beta_1\cS^{-1}+\beta_2\cS^{-2}+\ldots\,.
\]
The coefficients $\alpha_k,\beta_k\in\cF_Q$ can be found recursively:
\begin{equation}\label{Phi_alphabeta}\begin{array}{ll}
\alpha_0=\cS^{-1}\left(\frac{Q_{u_{1,0}}}{Q_{u_{1,1}}}\right),&
\beta_0=\cS^{-1}\left(\frac{Q_{u_{1,1}}}{Q_{u_{1,0}}}\right),    \\
\alpha_1=\cS^{-1}\left(\frac{Q_{u_{0,0}}}{Q_{u_{1,1}}}-
\frac{Q_{u_{0,1}}}{Q_{u_{1,1}}}\, \cS^{-1}\left(\frac{Q_{u_{1,0}}}{Q_{u_{1,1}}}\right)\right),&
\beta_1=\cS^{-1}\left(\frac{Q_{u_{0,1}}}{Q_{u_{1,0}}}-
\frac{Q_{u_{0,0}}}{Q_{u_{1,0}}}\, \cS^{-1}\left(\frac{Q_{u_{1,1}}}{Q_{u_{1,0}}}\right)\right),\\
 \alpha_{n+1}=(-1)^n\cS^{-1}\left(\frac{Q_{u_{0,1}}}{Q_{u_{1,1}}}\, \alpha_n\right),&
\beta_{n+1}=(-1)^n\cS^{-1}\left(\frac{Q_{u_{0,0}}}{Q_{u_{1,0}}}\, \beta_n\right),\quad n\ge 1.
 \end{array}
\end{equation}
Then (\ref{cond0}) follows  from (\ref{sigma0}) and formulas (\ref{cond1}) and
(\ref{cond2}) follow from  (\ref{Rnconslaws}) and (\ref{sigma_n}) for  $n=1$ and
$n=2$ respectively.\hfill$\blacksquare$

There is no  obstacle to compute equations corresponding to $n=3,4,5, $ etc. Of
course, the expressions get bigger. Obviously, if we consider the recursion
operators $\hat{\gR}$ in the $\cT$ direction, we immediately get :
\begin{Pro}
If a recursion operator $\hat{\gR}$ is represented by a first order formal Laurent
$\cT$ series $\hat{\gR}_L=\hat{r}_1 \cT+\hat{r}_0+\hat{r}_{-1}\cT^{-1}+\cdots$, then
\begin{eqnarray}\label{cond0T}
({\rm i})&\qquad & (\cS-{\bf 1})(\ln \hat{r}_1)=(\cT-{\bf
1})\cT^{-1}\left(\ln\frac{Q_{u_{1,1}}}{Q_{u_{0,1}}}\right),\\ \label{cond1T}
({\rm ii})&\qquad &  (\cS-{\bf 1})( \hat{r}_0)=(\cT-{\bf 1})\cT^{-1}\left(\hat{r}_1
\hat{F}
\right),
\\ \label{cond2T}
({\rm iii})&\qquad &(\cS-{\bf 1})(
\hat{r}_{-1}\cS^{-1}(\hat{r}_1)+\hat{r}_0^2+\hat{r}_1\cS(\hat{r}_{-1}))=(\cT-{\bf 1})
(\hat{\sigma}_2),
\end{eqnarray}
where
\[
\hat{\sigma}_2 \,= \, \cS^{-1}(\hat{r}_1\, \hat{F})\,\left\{ \cS^{-1}(\hat{r}_0)\,
+\, \hat{r}_0\, -\,\cS^{-2}\left( \hat{r}_1 \,\hat{F}\right)\right\}\, -\, (
1+\cS^{-1}) \left( \hat{r}_1 \,\hat{G}\, \cS^{-1} \left(\hat{r}_1 \, \hat{F}\right)
\right)\,,
\]
and $\hat{F},\hat{G}$ denote
\[
\hat{F}=\frac{Q_{u_{1,0}} \cT^{-1}(Q_{u_{0,1}})-
Q_{u_{0,0}}\cT^{-1}(Q_{u_{1,1}})}{Q_{u_{0,1}}\cT^{-1}(Q_{u_{1,1}})},\qquad
\hat{G}=\frac{Q_{u_{0,0}}}{Q_{u_{0,1}}}.
\]
\end{Pro}

It is not difficult to show that the existence of a formal series satisfying equation
(\ref{recursion_eq}) is the necessary condition for the existence of a hierarchy of
higher symmetries (of increasing order) for a difference equation. Thus, formulas
(\ref{cond0}--\ref{cond2}) and (\ref{cond0T}--\ref{cond2T}) can be seen as
integrability conditions for the  difference equation defined by function $Q$.
Indeed, the right hand sides of these formulas are expressed in the terms of the
difference equation only, and it is a rather non-trivial fact that the corresponding
left hand sides are in the image space of the operators $\cT-{\bf 1}$ or  $\cS-{\bf
1}$ respectively. If this is the case we can find the coefficients $r_1, \hat{r}_1$
and proceed to the next conditions (\ref{cond1}) and (\ref{cond1T}) and so on.

Theorem \ref{The4} provides infinitely many integrability condition of that type.
Integrability conditions (\ref{cond0}) and (\ref{cond0T}) have been earlier found by
Levi and Yamilov \cite{LeviYami2009} from the symmetry analysis of difference
equations. They actually found two more conditions of this type, which in our setup
correspond to the Taylor formal series  (see footnote \ref{footnote1}) and can be
easily recovered. Using the Taylor formal series we can double the number of the
conditions, which would be appropriate for asymmetric or linearisable equations.

It is not a straightforward exercise to apply the integrability conditions obtained. Nevertheless, they can be perfectly used for testing integrability as well as for classification of integrable difference equations (see for example \cite{LeviYami2009}).

\section{Recursion operator, symmetries and conservation laws for the Viallet equation}

In this section we apply the theory developed in the previous sections to the case of
the Viallet equation, as well as to the ABS equations. The first is characterized as
integrable because its algebraic entropy vanishes, \cite{Viallet}, while the
multidimensional consistency of latter implies their integrability, \cite{ABS}. Here,
we will show that they are integrable also in the sense that they admit infinite
hierarchies of symmetries by constructing recursion operators for all of them. In
order to make our presentation self-contained, we first introduce some properties of
these equations. Next we present two symmetries and a recursion operator for the
Viallet equation and demonstrate how it is related to corresponding operators for all
of the ABS equations through an explicit example. Finally, we present some higher
order conservation laws for all of the equations under consideration.

Throughout this section we assume that, the defining function $Q$ of equation
(\ref{QV}) is irreducible. This means that parameters $a_i$ cannot take values for
which $Q$ can be factorized as a product of two polynomials. By making  appropriate
choices of the coefficients $a_i$ we can reduce the Viallet equation to every
equation from the ABS list (see Appendix). The ABS equations depend on two so-called
lattice parameters $\alpha$ and $\beta$.

Let us introduce two polynomials in terms of $Q$ for equation (\ref{QV}), which will
be used in the rest of this section. They are determined by the discriminants of $Q$,
\begin{eqnarray}
h(u_{0,0},u_{1,0}) &:=&  Q\, \partial_{u_{0,1}} \partial_{u_{1,1}} Q\, -\,
\partial_{u_{0,1}} Q \, \partial_{u_{1,1}} Q \,, \\
\hat{h}(u_{0,0},u_{0,1}) &:=&  Q\, \partial_{u_{1,0}} \partial_{u_{1,1}} Q\, -\,
\partial_{u_{1,0}} Q \, \partial_{u_{1,1}} Q \,,\label{h-polynomials-def}.
\end{eqnarray}
Polynomials $h(u_{0,0},u_{1,0}),\hat{h}(u_{0,0},u_{0,1})$ are symmetric and
biquadratic \cite{ABS1}. In the case of  the ABS equations the polynomial $h$ can be
factorised and presented in the form
\begin{equation}
h(u,x) \,=\, k(\alpha,\beta)\, f(u,x;\alpha)\,, \label{ABS-h}
\end{equation}
where $k$ is a skew-symmetric function of the lattice parameters and  $f$ is
symmetric and biquadratic polynomial of $u$ and $x$ \cite{ABS}. Polynomials
$f(u,x;\alpha)$ for the ABS equations are listed in the Appendix.

The covariance of these equations is another useful property for our presentation.
For the Viallet equation this means that it is invariant under interchanging $u_{1,0}$
and $u_{0,1}$. For all of the ABS equations, covariance means that we have to
interchange not only $u_{1,0}$ and $u_{0,1}$, but the lattice parameters $\alpha$ and
$\beta$ as well. This property requires us to study symmetries, recursion operators and
conservation laws only for the one direction of the lattice (and recover the
complementary set using the covariance). We shall assume that symmetries and
canonical densities are elements of $\cF_\bs$ and the recursion operator is a
$\bs$-pseudo-difference operator. In most cases functions shifted by $\cS$ only will
contribute to the expressions. For this reason from now on, we will use one-index
notation for the shifts of the  polynomial $h$, i.e.
$$h \,=\, h(u_{0,0},u_{1,0})\,,\quad h_{i}\,=\,\cS^i h(u_{0,0},u_{1,0})\,,$$
and symmetries, e.g. $K^{(1)}_{j}=\cS^j ( K^{(1)})$. We shall omit the index zero for
unshifted functions.

We now turn our attention to the symmetry analysis of the equations. It is known
\cite{X,TTX} that these equations admit a generalised symmetry $K^{(1)}$ of order
$(-1,1)$, which is given by the formula
\begin{equation}
K^{(1)} \,:=\, \frac{h}{u_{1,0}-u_{-1,0}} - \frac{1}{2} \partial_{u_{1,0}} h\, = \,
\frac{h_{-1}}{u_{1,0}-u_{-1,0}} + \frac{1}{2} \partial_{u_{-1,0}} h_{-1}\,. \label{K1}
\end{equation}

Using Definition \ref{DefSymmetry} and formula (\ref{sym_def}), we can directly
verify the following result:
\begin{Pro} The Viallet equation (\ref{QV}) possesses a symmetry of order $(-2,2)$ which
has the form
\begin{eqnarray}
K^{(2)} \,=\, \frac{h\,h_{-1}}{(u_{1,0}-u_{-1,0})^2}\left( \frac{1}{u_{2,0}-u_{0,0}} \,+\,
\frac{1}{u_{0,0}-u_{-2,0}} \right) \,.\label{K2}
\end{eqnarray}
\end{Pro}

Actually, the Viallet equation possesses infinitely many local generalised
symmetries. These hierarchies of symmetries can be constructed by applying
successively a recursion operator $\gR$ on the {\em seed } or {\em root} symmetries
$K^{(1)}$ and $K^{(2)}$, the starting points for a hierarchy of symmetries
\cite{mr1974732}.

\begin{The}\label{reVia}
The Viallet equation (\ref{QV}) possesses a recursion operator $\gR=\cH\circ \J$, where
operators $\cH$ and $\J$ are defined as
\begin{eqnarray}
\J &=&\frac{1}{h} \cS - \cS^{-1} \frac{1}{h}\ ; \label{symplectic}\\
{\cal{H}} &=& \frac{h_{-1}\, h\, h_{1}}{(u_{1,0}-u_{-1,0})^2 (u_{2,0}-u_{0,0})^2}\, {\cal{S}}\,-\,
{\cal{S}}^{-1} \frac{h_{-1}\, h\, h_{1}}{(u_{1,0}-u_{-1,0})^2 (u_{2,0}-u_{0,0})^2}  \nonumber \\
& & \nonumber \\
& & +\, 2\,K^{(1)}\, {\cal{S}} \, ({\cal{S}}-{\bf 1})^{-1}\circ K^{(2)}\, +\, 2\, K^{(2)}\,
({\cal{S}}-{\bf 1})^{-1}\circ K^{(1)}\,. \label{hamilt-gen}
\end{eqnarray}
\end{The}

In this Theorem  $\cH$  and $\J$ are compatible Hamiltonian and symplectic operators
for the Viallet equation, they will be discussed in details in our paper \cite{MWX2}
including a complete proof of the Theorem. In the same paper, we will also show that
both $\gR^\ell (K^{(1)})$ and $\gR^\ell (K^{(2)})$ are in $\cF_{\bs}$ for all
$\ell\in\N$.  This is not obvious at all, since the Hamiltonian operator $\cH$  
contains a pseudo-difference operator $({\cal{S}}\,-\,{\bf 1})^{-1} $.
Symmetries $K^{(1)}$ and $K^{(2)}$ are called the seed (or root) symmetries for the
Viallet equation. Starting from them, two infinite hierarchies of symmetries can be
generated by applying the operator $\gR^\ell$. Later in this Section  we shall
demonstrate the proof of Theorem \ref{reVia} for the H1 equation (Theorem
\ref{recH1The}).

As an application, we now compute explicitly the action of $\gR$ on the first
symmetry $K^{(1)}$.
\begin{Pro} The Viallet equation (\ref{QV}) possesses a symmetry of order $(-3,3)$ which
is given by
\begin{eqnarray}
 K^{(3)}\,:=\,\gR (K^{(1)}) &=&  \frac{h\, h_{-1}}{(u_{1,0}-u_{-1,0})^2}
\left(\frac{K^{(1)}_{2}}{(u_{2,0}-u_{0,0})^2}+ \frac{K^{(1)}_{-2}}{(u_{0,0}-u_{-2,0})^2} \right) \nonumber \\
 & & +\left(\frac{1}{u_{2,0}-u_{0,0}} + \frac{1}{u_{0,0}-u_{-2,0}} \right) K^{(1)} K^{(2)} \,.\label{K3}
\end{eqnarray}
\end{Pro}
{\bf Proof}. To shorten the expression, we introduce the notation:
$$w \,:=\,\frac{1}{u_{1,0}-u_{-1,0}}\,, \quad w_k \,=\cS^k(w)\,. $$
First we work out the product of operator $\cH$ and $\J$
\begin{eqnarray}
&&\gR = h\, h_{-1}\,w^2
\left(w_{1}^2 \cS^2 + \cS^{-2}w_{1}^2 \right)
+2 K^{(1)} K^{(2)} \left(\frac{1}{h} \cS+\cS^{-1}\frac{1}{h}\right) \nonumber \\
&&\quad  - w^2 \left(h_{-1}\, h_{1} w_{1}^2+ h_{-2}\, h\,w_{-1}^2\right)
+\frac{2}{h_{-1}} \left( K^{(1)} K^{(2)}_{-1} + K^{(2)} K^{(1)}_{-1} \right) \nonumber \\
& &\quad +\, 2\,\,K^{(1)}\, (\cS-{\bf 1})^{-1}\circ
\left(\frac{K^{(2)}_{-1}}{h_{-1}}\,-\frac{K^{(2)}_{1}}{h} \right) +\,
2\,\,K^{(2)}\, (\cS-{\bf 1})^{-1}\circ 
\left(\frac{K^{(1)}_{-1}}{h_{-1}}\,-\frac{K^{(1)}_{1}}{h} \right)\, .
\label{weakR}
\end{eqnarray}
Acting with operator $\gR$ on the symmetry $K^{(1)}$ and using the identities
\begin{eqnarray*}
K^{(i)} {\cal{I}} K^{(i)} &=& (\cS-{\bf 1}) \left(\frac{1}{h_{-1}}\, K^{(i)} \, K^{(i)}_{-1} \right)\,,\quad i\,=\,1,\,2\ ;\\
 \\
K^{(2)} {\cal{I}} K^{(1)} &=& (\cS-{\bf 1})\left(\frac{-1}{h} K^{(2)} \, K^{(1)}_1 +
\frac{1}{2} \left(\frac{1}{u_{2,0}-u_{0,0}}+ \frac{1}{u_{0,0}-u_{-2,0}} \right) K^{(2)} \right)  \\
& &+ \,(\cS-{\bf 1}) (\cS+{\bf 1}) \left(\frac{h\, h_{-2}}{2 (u_{1,0}-u_{-1,0})^2 (u_{0,0}-u_{-2,0})^2} \right)\ ;\\
K^{(1)}\, {\cal{I}} \, K^{(2)} &=& -\, K^{(2)}\,{\cal{I}}\,K^{(1)} \,+\, (\cS-{\bf 1})
\left(\frac{1}{h_{-1}}K^{(1)} \, K^{(2)}_{-1} +\frac{1}{h_{-1}} K^{(2)} \, K^{(1)}_{-1} \right);
\end{eqnarray*}
we obtain expression (\ref{K3}) for  symmetry $K^{(3)}$.
\hfill $\blacksquare$

Using the above identities, one can easily compute the symmetry $K^{(4)}=\gR
(K^{(2)})$ of order $(-4,4)$. Applying $\gR^N$ to the seeds $K^{(1)}$ and $K^{(2)}$
we can generate two hierarchies of symmetries of order $(-2N-1,2N+1)$ and
$(-2N-2,2N+2)$ respectively for the Viallet equation.

\begin{Rem} \label{ConneQVABS}
Symmetries and recursion operators for the ABS equations can be obtained from the
symmetries of the Viallet equation and the operator $\gR=\cH\J$, given in Theorem
\ref{reVia}, by replacing the polynomial $h$ (everywhere in the symmetries and the
operators $\J$, $\cH$) with the corresponding polynomials $f :=
f(u_{0,0},u_{1,0},\alpha)$. For all ABS equations the polynomials $f$ are listed in
the Appendix. Equations H1--H3, Q1 and Q3$_{\delta=0}$ admit point symmetries
\cite{TTX}. For these equations, we find additional Hamiltonian operators of zero
Laurent order which are given in the Appendix.
\end{Rem}

\begin{Pro} If the recursion operator $\gR$, defined in Theorem \ref{reVia},
is restricted to the H1 equation
\begin{eqnarray}\label{H1re}
(u_{0,0}-u_{1,1})(u_{1,0}-u_{0,1}) = \alpha - \beta , \end{eqnarray}
 then it
becomes
$$\gR \, = \,\gR_{H1}^2\,,$$
where
\begin{eqnarray}\label{Q5-H1H}
\gR_{H1} \,= \frac{1}{u_{1,0}-u_{-1,0}}(\cS+{\bf 1})(\cS-{\bf 1})^{-1} \frac{1}{u_{1,0}-u_{-1,0}}
(\cS-\cS^{-1}) \, .
\end{eqnarray}
\end{Pro}
{\bf Proof}. The restriction to the H1 equation implies that  $h = 1$. Thus, the
first two symmetries (\ref{K1}) and (\ref{K2}) become
$$ K^{(1)} \,=\, \frac{1}{u_{1,0}-u_{-1,0}} \quad {\mbox{and}} \quad  K^{(2)} \,=\,
\frac{1}{(u_{1,0}-u_{-1,0})^2} \left(\frac{1}{u_{2,0}-u_{0,0}}+\frac{1}{u_{0,0}-u_{-2,0}} \right) \,,$$
respectively. Substituting them into (\ref{weakR}), a direct calculation leads to the statement.
\hfill $\blacksquare$

The operator given by (\ref{Q5-H1H}) is a recursion operator of the H1 equation. To
prove it, we will use the following result.
\begin{Lem}  \label{def-gp}
Operator
\begin{eqnarray} \label{gp}
\gM &=& \frac{1}{(u_{2,1}-u_{0,1})^2} \cS + \frac{1}{(u_{1,0}-u_{-1,0})^2} \cS^{-1} +
\frac{2}{(u_{1,1}-u_{-1,1}) (u_{2,1}-u_{0,1})}\nonumber \\
& & + (\alpha-\beta) \left(\frac{1}{(u_{1,1}-u_{-1,1})^2(u_{1,0}-u_{0,1})^2} -
\frac{1}{(u_{2,1}-u_{0,1})^2(u_{2,0}-u_{1,1})^2} \right) \nonumber \\
& & - 2 \left(\frac{1}{u_{1,0}-u_{-1,0}} + \frac{\alpha-\beta}{(u_{2,0}-u_{0,0})(u_{1,0}-u_{0,1})^2} \right) \times \nonumber \\
& & \qquad \qquad (\cS-{\bf 1})^{-1} \circ\left( \frac{1}{u_{2,0}-u_{0,0}} -
\frac{u_{1,0}-u_{0,1}}{(u_{1,0}-u_{-1,0}) (u_{0,0}-u_{-1,1})} \right)
\end{eqnarray}
satisfies the following identity for all solutions of the H1 equation (\ref{H1re}):
\begin{eqnarray}
&&\left({\bf 1}\,+\,\frac{\alpha-\beta}{(u_{1,0}-u_{0,1})^2} \, \cS \right)\circ\gR_{H1}\,=\,\gM\circ\left({\bf 1}\,+
\,\frac{\alpha-\beta}{(u_{1,0}-u_{0,1})^2}\, \cS\right);\label{hatp}\\
&&\left(\cS\,+\,\frac{\alpha-\beta}{(u_{1,0}-u_{0,1})^2}\right)\circ \cT(\gR_{H1})
=\gM\circ \left(\cS\,+\,\frac{\alpha-\beta}{(u_{1,0}-u_{0,1})^2}\right).\label{hatp2}
\end{eqnarray}
\end{Lem}
{\bf Proof.}  Writing out explicitly both sides of equation (\ref{hatp}), we
factorise terms in both sides according to the different powers of $\cS$. Collecting
the coefficients of the different powers of $\cS$ in both sides of the resulting
expression, their equality follows by putting them into $\cF_0$. In the same way, we
can prove identity (\ref{hatp2}).  \hfill $\blacksquare$

\begin{The}\label{recH1The}
Operator $\gR_{H1}$ given by (\ref{Q5-H1H}) is a recursion operator of the H1
equation (\ref{H1re}) with
\begin{equation}\label{gp1}
\gP \,= \, (u_{1,0}-u_{0,1}) \,\gM \circ(u_{1,0}-u_{0,1})^{-1}\,.
\end{equation}
\end{The}
{\bf Proof. } From Theorem \ref{The2} it follows that, $\gR_{H1}$ is a recursion
operator of the H1 equation provided both formulas (\ref{recursion_eq}) and
(\ref{gP1}) are valid for operator $\gP$. We rewrite them as
\begin{eqnarray}
&&(Q_{u_{1,0}}\cS+Q_{u_{0,0}})\circ  \gR_{H1} =\gP\circ (Q_{u_{1,0}}\cS+Q_{u_{0,0}});\label{con1}\\
&&(Q_{u_{1,1}}\cS+Q_{u_{0,1}})\circ  \cT(\gR_{H1}) =\gP\circ (Q_{u_{1,1}}\cS+Q_{u_{0,1}}).\label{con2}
\end{eqnarray}
For the H1 equation, we have
\begin{eqnarray*}
&&Q_{u_{1,0}}\,\cS\,+\,Q_{u_{0,0}}\,=\,(u_{0,0}-u_{1,1})\,\cS\,+\,(u_{1,0}-u_{0,1})\,
=\,(u_{1,0}-u_{0,1})\left({\bf 1}\,+\,\frac{\alpha-\beta}{(u_{1,0}-u_{0,1})^2}\, \cS \right);\\
&&Q_{u_{1,1}}\,\cS\,+\,Q_{u_{0,1}}\,=\,-\,(u_{1,0}-u_{0,1})\,\cS\,-\,(u_{0,0}-u_{1,1})
=-\,(u_{1,0}-u_{0,1})\left(\cS\,+\,\frac{\alpha-\beta}{(u_{1,0}-u_{0,1})^2}\right).
\end{eqnarray*}
Using Lemma \ref{def-gp}, we can easily obtain (\ref{con1}) and (\ref{con2}) and thus
we proved the statement. \hfill $\blacksquare$

Having constructed a recursion operator for equation (\ref{QV}), one can use the
results of Section \ref{Rec&CL} to find conserved densities for this equation. 
If we apply directly Theorem \ref{The4} to the recursion operator (\ref{weakR}), since the Laurent order of the recursion operator
$\gR$ is two, we will get 
only the ``even'' part of the infinite sequence of the canonical conservation laws . Although  it is impossible to compute a square root of a generic difference
formal series,  for the recursion operator (\ref{weakR}) it can be done and thus
one can find a first order formal recursion operator
\[ \tilde{\gR}=\frac{h_{-1}}{(u_{1,0}-u_{-1,0})^2}\cS+\frac{2 K^{(1)}}{u_{0,0}-u_{-2,0}}\,-\,\partial_{u_{0,0}} K^{(1)}+\cdots
 \ ,
\]
such that $\tilde{\gR}^2=\gR$. The first order recursion operator provides us with a complete set of 
canonical conservation laws and the first three densities of this sequence are 
\begin{eqnarray}
\rho_0 &=& \ln \left( \partial_{u_{1,0}} K^{(1)} \right)\,=\, \ln h_{-1} \,-\, 2\, \ln
\left(u_{1,0}-u_{-1,0} \right), \\
\rho_1 &=& \frac{2 K^{(1)}}{u_{0,0}-u_{-2,0}}\,-\,\partial_{u_{0,0}} K^{(1)}, \\
\rho_2 &=&  \frac{-1}{(u_{1,0}-u_{-1,0})^2}
\left(\frac{h_{-1}\,h_{1}}{(u_{2,0}-u_{0,0})^2}+ \frac{h_{-2}\,
h}{(u_{0,0}-u_{-2,0})^2}\right) +\frac{2}{h_{-1}} \left( K^{(1)} K^{(2)}_{-1} +
K^{(2)} K^{(1)}_{-1} \right).
\end{eqnarray}
The first and the second densities are the logarithmic residue and the residue of  $\tilde{\gR}$, respectively, while $\rho_2$ is the residue of the recursion operator $\gR=\tilde{\gR}^2$ (\ref{weakR}).

According to Proposition \ref{PropIntegrCon1}, the
corresponding fluxes are
\begin{eqnarray}
\sigma_0 &=& \cS^{-1}\left(\ln\frac{Q_{u_{1,1}}}{Q_{u_{1,0}}}\right)\,,\\
\sigma_1 &=& \cS^{-1}\left(\frac{h_{-1}}{\left(u_{1,0}-u_{-1,0} \right)^2}\,F\right)\,,\\
\sigma_2 &=& \cS^{-1}\left(\frac{2\, F\,K^{(1)}\, K^{(2)}}{h}\right)
-\cS^{-2}\left( \frac{F\,h h_{-1}}{(u_{2,0}-u_{0,0})^2(u_{1,0}-u_{-1,0})^2}\right) \nonumber \\
& &- (1+\cS^{-1}) \left(\frac{Q_{u_{0,0}}}{Q_{u_{1,0}}}
\frac{h_{-1} h_{-2}}{(u_{1,0}-u_{-1,0})^2(u_{0,0}-u_{-2,0})^2} \cS^{-1}F\right)\,,
\end{eqnarray}
where
$$F \,=\,\frac{Q_{u_{0,1}}
 \,\cS^{-1}(Q_{u_{1,0}})}{Q_{u_{1,0}}\,\cS^{-1}(Q_{u_{1,1}})}\,-\,\frac{Q_{u_{0,0}}}{Q_{u_{1,0}}}\,, $$
and $Q$ is the defining polynomial of equation (\ref{QV}) (or the ABS equations).

The covariance of the equation allows us to produce three more conservation laws simply by
changing $u_{i,0}$ to $u_{0,i}$, $h$ to $\hat{h}$ and operator $\cS$ to $\cT$ in the above
densities and fluxes. On the other hand, these conservation laws can also be restricted to
the ABS equations in the way described in Remark \ref{ConneQVABS}.

For the H1 equation, its recursion operator (\ref{Q5-H1H}) is first order and we can
use relations (\ref{ResiduesOrder1}) to derive some conserved densities. Actually,
the following formulas for conserved densities are valid for the equations admitting
 Hamiltonian operators of zero Laurent order, i.e. equations H1--H3, Q1 and Q3$_{\delta=0}$. The conserved densities for all these equations can be written in a relatevly compact form (they
are equivalent to the ones given in Proposition \ref{PropIntegrCon1}):
\begin{eqnarray*}
\varrho_0&=&\ln (\cS^{-1}f)-2\ln (u_{1,0}-u_{-1,0}),\\
\varrho_1&=&\frac{f}{(u_{1,0}-u_{-1,0})(u_{2,0}-u_{0,0})×},\\
\varrho_2&=& (\varrho_1+\cS(\varrho_1))^2-\cS(\varrho_1)^2,\\
\varrho_3&=&  (\varrho_1+\cS(\varrho_1))^3-\cS(\varrho_1)^3+3\varrho_1\cS(\varrho_1)\cS^2(\varrho_1),\\
\varrho_4&=&
(\varrho_1+\cS(\varrho_1))^4-\cS(\varrho_1)^4+4\varrho_1\cS(\varrho_1)\cS^2(\varrho_1)(\varrho_1+2\cS(\varrho_1)
+\cS^2(\varrho_1)+\cS^3(\varrho_1)).
\end{eqnarray*}
For the H1 equation (\ref{H1re}), some of the above densities were found in
\cite{RasinSchiff} by the Gardner method.

\section{Conclusions}
The recursion operators presented in this paper produce infinite hierarchies of
symmetries and conservation laws for all ABS equations. For the H1 equation we have shown
that the proposed recursion operator, $\gR_{H_1}$, does satisfy defining equation
(\ref{rec}) and thus acting on its symmetry $K^{(1)}$ it will produce its new
symmetry $\gR_{H_1}(K^{(1)})$, provided such action is well defined (i.e. the result
belongs to $\cF_Q$). The proof that we can repeatedly apply the recursion operator
and each time  get a symmetry requires us to investigate the property of Nijenhuis
operators \cite{GD79,mr82g:58039} and to develop a construction similar to the famous
Lenard scheme for the Korteweg-de Vries equation \cite{ps05,wang09}.  The
generalisation of Lenard's scheme for difference equation as well as a complete proof
that the operator $\gR$ defined in the Theorem \ref{reVia} does satisfy (\ref{rec}) for the
Viallet equation will be published in our paper \cite{MWX2}.

For a quadrilateral equation the first integrability condition has been found by Levi
and Yamilov \cite{LeviYami2009}. In this paper we have proposed an infinite sequence of
integrability conditions in the form of canonical conservation laws. The Levi-Yamilov
condition is the first element of this sequence. We have shown that canonical conservation laws 
are necessary conditions for the existence of a formal recursion
operator for a difference equation. As in the case of partial differential
equations, cf. \cite{mr86i:58070, mr89e:58062, mr93b:58070}, one can
prove that the existence of a formal recursion operator follows from
the existence of an infinite hierarchy of symmetries of increasing order for a
difference equation (a general and detailed proof of this assertion will be published
elsewhere). There are no doubts that integrability conditions can be successfully used
 for testing and classification of integrable difference equations (see for example \cite{LeviYami2009}). 

Our theory can be  extended to difference equations of ``higher order'', i.e.
equations on a rectangular with function $Q$ depending on the set of variables
$\{u_{p,q}\,|\, 0\le p\le N\,,\ \, 0\le q\le M\}$. For the existence of the
elimination map $\cE$ we should assume that equation $Q=0$ can be uniquely resolved
with  respect to the corner variables $ u_{0,0},u_{N,0},u_{0,M}$ and $u_{N,M}$. In
this case the dynamical variables can be defined as
\[
 U_\bs=\{u_{n,p}\,|\, n\in\Z,\ 0\le p<M\},\quad  U_\bt=\{u_{q,m}\,|\, m\in\Z,\
 0\le q<N\},\quad U_0=U_\bs\cup U_\bt.
\]
We are working on the extension of our theory to the higer order case, the vector case, as well as to the
case when $Q$ is not assumed to be an affine-linear polynomial.

\section*{Acknowledgments}
We are grateful to V. E. Adler for his clarification about the Viallet equation. 
Also we would like to thank C. Athorne and A. Pillay for useful discussions.
P.X. is supported by the {\emph{Newton International Fellowship}} grant NF082473 entitled
``Symmetries and integrability of lattice equations and related partial differential equations''.
A.V.M. and P.X. would like to thank the University of Kent for its hospitality during their visits.

\section*{Appendix : The ABS equations}

For every equation in the ABS classification \cite{ABS} we present its defining polynomial $Q$,
the corresponding  choices for $a_i$ in (\ref{QV}), the corresponding polynomial $f$  and factor $k$ (\ref{ABS-h}),
the Hamiltonian operator $\cH$ and the seed (root) symmetry of a hierarchy.

The recursion operator for all of the ABS equations can be written as $\gR=\cH \circ\J$,
where the symplectic operator $\J$ is
\[ \J \,=\,\frac{1}{f}\, \cS \,- \,\cS^{-1}\, \frac{1}{f}\,,\]
and the Hamiltonian operator $\cH$ is either
\begin{eqnarray} \label{hamop}
{\cal{H}} &=& \frac{f_{-1}\, f\, f_{1}}{(u_{1,0}-u_{-1,0})^2
(u_{2,0}-u_{0,0})^2}\, {\cal{S}}\,-\,
{\cal{S}}^{-1} \frac{f_{-1}\, f\, f_{1}}{(u_{1,0}-u_{-1,0})^2
(u_{2,0}-u_{0,0})^2}  \nonumber \\
& & \nonumber \\
& & +\, 2\, \left(\,K^{(1)}\, {\cal{S}} \, (\cS-{\bf 1})^{-1} 
K^{(2)}\, +\, K^{(2)}\,
(\cS-{\bf 1})^{-1} K^{(1)}\, \right)\,,
\end{eqnarray}
which is valid for all of the ABS equations, or the  one given in the list for the
case of H1--H3, Q1 and Q3$_{\delta=0}$. In (\ref{hamop}) and in the lists, $K^{(1)}$
and $K^{(2)}$ denote the first symmetries of the ABS equations, which have the
following form
\begin{eqnarray*}
K^{(1)} &=& \frac{f}{u_{1,0}-u_{-1,0}} - \frac{1}{2}
\partial_{u_{1,0}} f\, = \,
\frac{f_{-1}}{u_{1,0}-u_{-1,0}} + \frac{1}{2} \partial_{u_{-1,0}} f_{-1}\,, \\
K^{(2)} &=& \frac{f\,f_{-1}}{(u_{1,0}-u_{-1,0})^2}\left(
\frac{1}{u_{2,0}-u_{0,0}} \,+\,
\frac{1}{u_{0,0}-u_{-2,0}} \right) \,.
\end{eqnarray*}

\begin{flushleft}
\begin{tabular}{ll}
& \\
{\bf{H1}} &  \\
Equation & $ Q = (u_{0,0}-u_{1,1})(u_{1,0}-u_{0,1}) - \alpha + \beta $\\
Parameters & $a_3 = - a_5 = 1$, $a_7 = \beta - \alpha$, $a_1=a_2=a_4=a_6=0$ \\
Polynomial $h$ & $f = 1$\\
 & $k = \beta-\alpha$ \\
$\cH$& $\cH \, = \, K^{(1)} (\cS+{\bf 1}) (\cS-{\bf 1})^{-1} K^{(1)} $ \\
Seed symmetry & $K^{(1)}$ \\
\end{tabular}

\begin{tabular}{ll}
& \\
{\bf{H2}} & \\
Equation & $ Q = (u_{0,0}-u_{1,1})(u_{1,0}-u_{0,1}) +(\beta-\alpha)
(u_{0,0}+u_{1,0}+u_{0,1}+u_{1,1}) - \alpha^2 + \beta^2$\\
Parameters & $a_3 = - a_5 = 1$, $a_6 = \beta - \alpha$, $a_7 = \beta^2
- \alpha^2$, $a_1=a_2=a_4=0$ \\
Polynomial $h$ & $f = 2 (u_{0,0} + u_{1,0} + \alpha)$ \\
 & $k = \beta-\alpha $ \\
$\cH$& $ \cH \,=\, K^{(1)} (\cS+{\bf 1}) (\cS-{\bf 1})^{-1} K^{(1)} - (\cS -
{\bf 1}) (\cS+{\bf 1})^{-1} $ \\
Seed symmetry & $K^{(1)}$ \\
\end{tabular}

\begin{tabular}{ll}
& \\
{\bf{H3}} & \\
Equation & $Q =  \alpha (u_{0,0} u_{1,0}+u_{0,1} u_{1,1}) - \beta
(u_{0,0} u_{0,1}+u_{1,0} u_{1,1}) + \delta (\alpha^2-\beta^2) $\\
Parameters & $a_3 =\alpha$, $ a_5 = -\beta$, $a_7 =
\delta(\alpha^2-\beta^2)$, $a_1=a_2=a_4=a_6=0$ \\
Polynomial $h$ & $f = u_{0,0} u_{1,0} + \alpha \delta$ \\
 & $k = \alpha^2 - \beta^2 $ \\
$\cH$& $ \cH \,=\,\ K^{(1)} (\cS+{\bf 1}) (\cS-{\bf 1})^{-1} K^{(1)} -
\frac{1}{4} u_{0,0} (\cS-{\bf 1}) (\cS+{\bf 1})^{-1} u_{0,0}$ \\
Seed symmetry & $K^{(1)}$ \\
\end{tabular}

\begin{tabular}{ll}
& \\
{\bf{Q1}} & \\
Equation & $Q= \alpha (u_{0,0}-u_{0,1}) (u_{1,0}- u_{1,1}) - \beta
(u_{0,0}- u_{1,0}) (u_{0,1} -u_{1,1})  + \delta^2 \alpha \beta
(\alpha-\beta) $ \\
Parameters & $a_3 = \alpha$, $a_4 = \beta - \alpha$, $a_5 = -\beta$,
$a_7 = \delta^2 \alpha \beta (\alpha-\beta)$ \\
Polynomial $h$ & $f = ((u_{0,0}-u_{1,0})^2 - \alpha^2 \delta^2)/\alpha $ \\
 & $k = \alpha \beta (\beta-\alpha) $ \\
$\cH$ & $  \cH \,=\, K^{(1)} (\cS+{\bf 1}) (\cS-{\bf 1})^{-1} K^{(1)} -
\delta^2 (\cS+{\bf 1}) (\cS-{\bf 1})^{-1}  $ \\
Seed symmetry & $K^{(1)}$ \\
\end{tabular}

\begin{tabular}{ll}
& \\
{\bf{Q2}} & \\
Equation & $Q=\alpha (u_{0,0}-u_{0,1}) (u_{1,0}- u_{1,1}) - \beta
(u_{0,0}- u_{1,0}) (u_{0,1} -u_{1,1})$ \\
& $  + \alpha \beta (\alpha-\beta) (u_{0,0}+u_{1,0}+u_{0,1}+u_{1,1}) -
\alpha \beta(\alpha-\beta) (\alpha^2-\alpha \beta + \beta^2)$\\
Parameters & $a_3 = \alpha$, $a_4 = \beta - \alpha$, $a_5 = -\beta$,
$a_6 = \alpha \beta (\alpha-\beta)$,
$a_7 = \alpha \beta (\beta-\alpha) (\alpha^2-\alpha \beta + \beta^2)$ \\
Polynomial $h$ & $f = ((u_{0,0}-u_{1,0})^2 - 2 \alpha^2
(u_{0,0}+u_{1,0}) + \alpha^4)/\alpha$ \\
 & $k = \alpha \beta (\beta-\alpha) $ \\
$\cH$ & given by (\ref{hamop})\\
Seed symmetries & $K^{(1)}$, $K^{(2)}$\\
\end{tabular}

\begin{tabular}{ll}
& \\
{\bf{Q3}} & \\
Equation & $ Q=(\beta^2-\alpha^2) (u_{0,0} u_{1,1}+u_{1,0} u_{0,1}) +
\beta (\alpha^2-1) (u_{0,0} u_{1,0}+u_{0,1} u_{1,1})$ \\
& $- \alpha (\beta^2-1) (u_{0,0} u_{0,1}+u_{1,0} u_{1,1}) -
\frac{\delta^2 (\alpha^2-\beta^2) (\alpha^2-1) (\beta^2-1)}{4 \alpha
\beta}$\\
Parameters & $a_3 = \beta (\alpha^2-1)$, $a_4 = \beta^2-\alpha^2$,
$a_5 = - \alpha (\beta^2-1)$, $a_7 = - \frac{\delta^2
(\alpha^2-\beta^2) (\alpha^2-1) (\beta^2-1)}{4 \alpha \beta}$  \\
Polynomial $h$ & $f = \frac{1}{4\alpha(\alpha^2-1)} (4 \alpha
(\alpha u_{0,0}-u_{1,0}) (\alpha u_{1,0} - u_{0,0}) - (\alpha^2-1)^2
\delta^2) $ \\
& $k = (\alpha^2 - \beta^2)(\alpha^2 -1)(\beta^2 - 1) $\\
$\cH$ & (i.) $\delta = 0$: $\cH = K^{(1)}( {\cal{S}}+{\bf 1})
(\cS-{\bf 1})^{-1} K^{(1)} - \frac{1}{4} u_{0,0} ( {\cal{S}}+{\bf 1})
(\cS-{\bf 1})^{-1} u_{0,0}$\\
& (ii.)\  $\delta\ne 0$: given by (\ref{hamop})\\
Seed symmetries & (i.)\  $\delta = 0$: $K^{(1)}$; (ii.)\ $\delta \ne
0$: $K^{(1)}$, $K^{(2)}$
\end{tabular}

\begin{tabular}{ll}
 \\
{\bf{Q4}} & \\
Equation & $Q={\rm{sn}}(\alpha) (u_{0,0} u_{1,0}+u_{0,1} u_{1,1}) -
{\rm{sn}}(\beta) (u_{0,0} u_{0,1}+u_{1,0} u_{1,1}) $ \\
&$-{\rm{sn}}(\alpha-\beta) (u_{0,0} u_{1,1}+u_{1,0} u_{0,1})$
 $+ k\,{\rm{sn}}(\alpha)\, {\rm{sn}}(\beta) \,{\rm{sn}}(\alpha-\beta)
\,(1 + u_{0,0} u_{1,0} u_{0,1} u_{1,1})$\\
Parameters & $a_1 = a_7 = k\,{\rm{sn}}(\alpha)\, {\rm{sn}}(\beta)
\,{\rm{sn}}(\alpha-\beta)$, $a_3 = {\rm{sn}}(\alpha)$, $a_4 =
-{\rm{sn}}(\alpha-\beta)$, $a_5 = - {\rm{sn}}(\beta)$  \\
Polynomial $h$ & $f = -k\, {\rm{sn}}(\alpha) (1+u_{0,0}^2 u_{1,0}^2)
+ \frac{1}{{\rm{sn}}(\alpha)} \left( u_{0,0}^2 +u_{1,0}^2 - 2
{\rm{cn}}(\alpha) {\rm{dn}}(\alpha) u_{0,0} u_{1,0} \right)$ \\
& $k = {\rm{sn}}(\alpha) {\rm{sn}}(\beta) {\rm{sn}}(\beta-\alpha)$\\
$\cH$ & given by (\ref{hamop})\\
Seed symmetries & $K^{(1)}$, $K^{(2)}$ \\
& \\
\end{tabular}
\end{flushleft}


\begin{thebibliography}{10}

\bibitem{mr91k:58005}
Zakharov, V. E. (ed.) :
\newblock {\em What is integrability?}
\newblock Springer-Verlag, Berlin (1991)

\bibitem{integrability}
Mikhailov, A. V. (ed.) :
\newblock {\em Integrability}.
\newblock Lecture Notes in Physics 767, Springer (2009)

\bibitem{mr86i:58070}
Sokolov, V. V., Shabat, A. B.:
\newblock Classification of integrable evolution equations.
\newblock In {\em Mathematical Physics Reviews, Vol. 4}, of {\em
  Soviet Sci. Rev. Sect. C: Math. Phys. Rev.}, pages 221--280. Harwood Academic
  Publ., Chur (1984)

\bibitem{mr89g:58092}
 Mikhailov, A. V., Shabat, A.~B., Yamilov, R. I.:
\newblock Extension of the module of invertible transformations.
  Classification of integrable systems.
\newblock {\em Comm. Math. Phys.} 115(1),1--19 (1988)

\bibitem{mr89e:58062}
Mikhailov, A.~V., Shabat,  A.~B., Yamilov, R.~I.:
\newblock Symmetry approach to the classification of nonlinear equations.
  Complete lists of integrable systems.
\newblock {\em Uspekhi Mat. Nauk} 42(4(256)), 3--53 (1987)

\bibitem{mr93b:58070}
Mikhailov, A. V, Shabat,  A.~B., Sokolov, V.~V.:
\newblock The symmetry approach to classification of integrable equations.
\newblock In {\em What is integrability?}, Springer Ser. Nonlinear Dynamics,
  pp 115--184. Springer, Berlin (1991)

\bibitem{mr99g:35058}
 Sanders, J.~A., Wang, J. P.:
\newblock On the integrability of homogeneous scalar evolution equations.
\newblock {\em J. Differential Equations} 147(2), 410--434 (1998)

\bibitem{wang98}
Wang, J. P.:
\newblock {\em Symmetries and Conservation Laws of Evolution Equations}.
\newblock PhD thesis, Vrije Universiteit/Thomas Stieltjes Institute, Amsterdam (1998)

\bibitem{asy}
 Adler, V.~E., Shabat, A.~B., Yamilov, R.~I.:
\newblock Symmetry approach to the integrability problem.
\newblock {\em Theor. Math. Phys.} 125, 1603--1661 (2000)

\bibitem{Yami1}
Yamilov, R.~I.:
\newblock Classification of discrete evolution equations.
\newblock {\em Upsekhi Mat. Nauk} 38, 155--156 (1983)

\bibitem{Yami}
Yamilov, R.~I.:
\newblock Symmetries as integrability criteria for differential difference
  equations.
\newblock {\em J. Phys. A: Math. Gen.} 39, R541--R623 (2006)

\bibitem{NC}
Nijhoff, F.~W., Capel, H.~W.:
\newblock The discrete Korteweg-de Vries equation.
\newblock {\em Acta Appl. Math.} 39, 133--158 (1995)

\bibitem{GHRV}
Grammaticos, B., Halburd, R.~G., Ramani, A., Viallet, C-M:
\newblock How to detect the integrability of discrete systems.
\newblock {\em J. Phys. A: Math. Theor.} 42, 454002 (2009)

\bibitem{NW}
Nijhoff, F. W., Walker, A. J.:
\newblock The discrete and continuous Painlev{\'e} hierarchy and the Garnier
  system.
\newblock {\em Glasgow Math. J.} 43A, 109�123 (2001)

\bibitem{BobSuris}
Bobenko, A.~I., Suris, Yu.~B.:
\newblock Integrable systems on quad-graphs.
\newblock {\em Int. Math. Res. Notices} 11, 573--611 (2002)

\bibitem{Nijhoff2002}
Nijhoff, F.~W.:
\newblock Lax pair for the Adler (lattice Krichever-Novikov) system.
\newblock {\em Phys. Lett. A} 297, 49--58 (2002)

\bibitem{ABS}
 Adler, V.~E., Bobenko, A.~I., Suris, Yu.~B.:
\newblock Classification of integrable equations on quad-graphs. The
  consistency approach.
\newblock {\em Comm. Math. Phys.} 233, 513--543 (2003)

\bibitem{ABS1}
Adler, V.~E., Bobenko, A.~I., Suris, Yu.~B.:
\newblock Discrete nonlinear hyperbolic equations. Classification of integrable
  cases.
\newblock {\em Funct. Anal. Appl.} 43, 3--21 (2009)

\bibitem{BellonViallet}
Bellon, M.~P., Viallet, C.~M.:
\newblock Algebraic entropy.
\newblock {\em Comm. Math. Phys.} 204, 425--437 (1999)

\bibitem{Viallet}
Viallet, C.:
\newblock Integrable lattice maps: $Q_V$, a rational version of $Q_4$.
\newblock {\em Glasgow Math. J.} 51A, 157--163 (2009)

\bibitem{X}
Xenitidis, P.:
\newblock Integrability and symmetries of difference equations: the
  Adler--Bobenko--Suris case.
\newblock In {\em Proceedings of the 4th Workshop ``Group Analysis of
  Differential Equations and Integrable Systems''} (2009)
\newblock arXiv: 0902.3954.

\bibitem{RHcons}
Rasin, O.~G., Hydon, P.~E.:
\newblock Conservation laws for integrable difference equations.
\newblock {\em J. Phys. A: Math. Theor.} 40, 12763--12773 (2007)

\bibitem{RasinSchiff}
Rasin, A., Schiff, J.:
\newblock Infinitely many conservation laws for the discrete KdV equation.
\newblock {\em J. Phys. A: Math. Theor.} 42, 175205 (2009)

\bibitem{rasin2010}
Rasin, A.:
\newblock Infinitely many symmetries and conservation laws for quad-graph
  equations via the Gardner method.
\newblock {\em arXiv:1001.0724} (2010)

\bibitem{LeviYami2009}
Levi, D., Yamilov, R. I.:
\newblock The generalized symmetry method for discrete equations.
\newblock {\em J. Phys. A: Math. Theor.} 42, 454012 (2009)

\bibitem{TTX}
Tongas, A., Tsoubelis, D., Xenitidis, P.:
\newblock Affine linear and $D_4$ symmetric lattice equations: symmetry
  analysis and reductions.
\newblock {\em J. Phys. A: Math. Theor.} 40, 13353--13384 (2007)

\bibitem{mr1845643}
Zhiber, A.~V., Sokolov, V.~V.:
\newblock Exactly integrable hyperbolic equations of Liouville type.
\newblock {\em Uspekhi Mat. Nauk} 56(1(337)), 63--106 (2001)

\bibitem{mr80i:58026}
Adler, M.:
\newblock On a trace functional for formal pseudo differential operators and
  the symplectic structure of the {K}orteweg-de {V}ries type
  equations.
\newblock {\em Invent. Math.} 50(3), 219--248 (1978/79)

\bibitem{mr1974732}
Sanders, J.~A., Wang, J. P. :
\newblock Integrable systems and their recursion operators.
\newblock {\em Nonlinear Anal.} 47, 5213--5240 (2001)

\bibitem{MWX2}
Mikhailov, A.V., Wang, J. P., Xenitidis, P.:
\newblock Extended {L}enard scheme for difference equations
\newblock {\em In preparation} 

\bibitem{GD79}
Gel'fand, I. M., Dorfman, I. Ya.:
\newblock Hamiltonian operators and algebraic structures related to them.
\newblock {\em Funct. Anal. Appl.} 13(4), 248--262 (1979)

\bibitem{mr82g:58039}
Fokas, A.~S., Fuchssteiner, B.:
\newblock On the structure of symplectic operators and hereditary symmetries.
\newblock {\em Lett. Nuovo Cimento (2)} 28(8), 299--303 (1980)

\bibitem{ps05}
Praught, J., Smirnov, G.:
\newblock Andrew {L}enard: A mystery unraveled.
\newblock {\em SIGMA} 1, 005 (2005)

\bibitem{wang09}
Wang, J. P.:
\newblock Lenard scheme for two-dimensional periodic Volterra chain.
\newblock {\em J. Math. Phys.} 50, 023506 (2009)

\end{thebibliography}
\end{document}